\documentclass[12pt,preprint]{aastex}
\shorttitle{{\sl HST} observations of UGC 4483}
\shortauthors{Izotov \& Thuan}
\begin{document}

\title{{\sl HST} observations of the cometary blue compact dwarf 
galaxy UGC 4483: 
a relatively young galaxy?{\footnote{Based on observations 
obtained with the NASA/ESA {\it{Hubble Space Telescope}} through the Space 
Telescope Science Institute, which is operated by AURA,Inc. under NASA
contract NAS5-26555.}}}
\author{Yuri I. Izotov}
\affil{Main Astronomical Observatory, National Academy of Sciences of Ukraine, 03680, Kyiv, Ukraine}
\email{izotov@mao.kiev.ua}

\and

\author{Trinh X. Thuan}
\affil{Astronomy Department, University of Virginia,
    Charlottesville, VA 22903}
\email{txt@virginia.edu}

\begin{abstract}
We present $V$ and $I$ photometry of the resolved stars in the 
cometary blue compact dwarf 
galaxy UGC 4483 using {\sl Hubble Space Telescope} Wide Field 
Planetary Camera 2 (WFPC2) images. The resulting $I$ vs. $(V-I)$
color-magnitude diagram (CMD) reaches limiting magnitudes $V$ =
27.5 mag and $I$ = 26.5 mag for photometric errors less than 0.2 mag. 
It reveals not only a young stellar population of blue main-sequence stars 
and blue and red
supergiants, but also an older evolved population of red giant and
asymptotic giant branch stars. The measured magnitude $I$ = 23.65 $\pm$ 0.10
mag of the red giant branch tip results in a distance modulus $(m-M)$ = 
27.63 $\pm$ 0.12, corresponding to a distance of 3.4 $\pm$ 0.2 Mpc.
The youngest stars are associated with the bright H {\sc ii} region at the
northern tip of the galaxy. The population of older stars is found
throughout the low-surface-brightness body of the galaxy and is
considerably more spread out than the young stellar population, suggesting 
stellar diffusion.
 The most striking 
characteristics of the CMD of UGC 4483 are the very blue colors of the
red giant stars and the high luminosity of the asymptotic
giant branch stars. Both of these characteristics are consistent with either: 
1) a very low metallicity ([Fe/H] = --2.4 like the most metal-deficient 
globular clusters) and an old age of 10 Gyr, or 
2) a higher metallicity ([Fe/H] = --1.4 as  derived
from the ionized gas emission lines) and a relatively young 
age of the oldest stellar population in UGC 4483, not exceeding $\sim$ 2 Gyr.
Thus our data do not exclude the possibility that UGC 4483 is a relatively 
young galaxy having formed its first stars only $\sim$ 2 Gyr ago.  
\end{abstract}

\keywords{galaxies: irregular --- galaxies: photometry --- 
galaxies: stellar content --- galaxies: distances and redshifts --- 
galaxies: evolution --- galaxies: individual (UGC 4483)}


\section{Introduction}

The dwarf irregular galaxy UGC 4483 can also be classified as a
cometary blue compact dwarf (BCD) galaxy in the 
morphological classification scheme of Loose \& Thuan (1986): 
the high-surface-brightness starburst region at the northern end of the galaxy 
represents the ``head'' of the comet while its
low-surface-brightness (LSB) elongated body represents its ``tail''. 
With equatorial coordinates $\alpha$(J2000) = 08$^{\rm h}$37$^{\rm m}$03\fs0,
$\delta$(J2000) = +69\arcdeg46\arcmin31\arcsec, it lies roughly midway between
M81 (7\fdg8 away) and NGC 2403 (7\fdg3 away) and likely belongs
to the M81 group of galaxies. Tikhonov \& Karachentsev (1993) have derived a 
distance to UGC 4483 of 3.6 Mpc from photometry of its brightest stars. 
For comparison, the Cepheid-determined distance to M81 is 3.47 Mpc and 
that to NGC 2403 is 2.82 Mpc. Recently
Dolphin et al. (2001) using {\sl Hubble Space Telescope} ({\sl HST})
observations have derived a lower distance of 3.2 Mpc and have argued that 
UGC 4483 belongs to the NGC 2403 subgroup rather than to the M81 subgroup.
From H {\sc i} 21 cm observations, Thuan \& Seitzer (1979) derived for
UGC 4483 a radial velocity of +156 km s$^{-1}$, a hydrogen mass 
$M$(H {\sc i}) = 4.1 $\times$ 10$^7$ $M_\sun$ (at a distance of 3.4 Mpc) 
and a gas mass fraction of 0.39. 

The first abundance determination of
the ionized gas surrounding the brightest H {\sc ii} region on the northern
side of the extended LSB body in UGC 4483 is by Skillman (1991) who derived an
oxygen abundance 12 + log O/H = 7.32. However, subsequent spectroscopic
studies have resulted in higher values: 7.52 (Skillman et al. 1994),
7.58 (Izotov, Thuan \& Lipovetsky 1994) and 7.54 ($Z_\odot$/23) 
(Izotov \& Thuan 1999). This low metallicity makes UGC 4483 a good candidate 
for being a relatively unevolved galaxy. Its nitrogen-to-oxygen
abundance ratio log N/O = --1.64 $\pm$ 0.03 falls in the narrow range of
values derived for other very metal-deficient BCDs with $Z$ $\leq$
$Z_\odot$/20 (Thuan, Izotov \& Lipovetsky 1995; Izotov \& Thuan 1999).
Izotov \& Thuan (1999) have argued that chemical abundances in BCDs can be
used to put constraints on their age. From the constancy of the C/O and N/O 
ratios as a function of O abundance, they have suggested 
that all galaxies with 
12 + log (O/H) $\la$ 7.6 ($Z_\odot$/20)
began to form stars less than $\sim$ 100 Myr ago. However, the most direct
way to derive the age of galactic stellar populations is by 
color-magnitude diagrams (CMD) of resolved stars and comparison with 
theoretical isochrones.

We report here {\sl HST} Wide Field and Planetary Camera 2 (WFPC2) $V$ and $I$ 
imaging of UGC 4483. These data are used to discuss the evolutionary status 
of this nearby dwarf galaxy. 
We describe the observations in Sect. 2. The distance to UGC 4483 is 
derived in Sect. 3. Its stellar populations are discussed in Sect. 4. 
In Sect. 5 we compare the CMD of UGC 4483 with those of other nearby irregular
and BCD galaxies available from the {\sl HST} archives. 
We summarize our findings in Sect. 6.

\section{Observations and data reduction}

  We obtained {\sl HST} images of UGC 4483 on 2000 August 12 during cycle 9
with the WFPC2 through filters F555W and F814W, which we will refer to 
hereafter as $V$ and $I$. The $V$ exposure was broken up in seven sub-exposures
and the $I$ filter in five sub-exposures to permit identification and removal 
of cosmic rays. The total exposure time was 9500s in $V$ and 6900s in $I$. 
The galaxy was positioned on the PC frame so that the high-surface-brightness
starburst region is at its North corner. The orientation of the WFPC2 has been
chosen so that the major axis of the galaxy lies along the diagonals of the 
PC and WF3 frames. Some parts of UGC 4483 were imaged also on the
WF2, WF3 and WF4 frames. The scale is 0\farcs046 per pixel for the PC frame 
and 0\farcs102 per pixel for the WF frames. 

   Preliminary processing of the raw images including 
flat-fielding was done at the Space Telescope Science Institute through the 
standard pipeline. Subsequent reductions were carried out at the Main
Astronomical Observatory of the Ukrainian Academy of Sciences
using IRAF{\footnote{IRAF is the Image Reduction and Analysis Facility
distributed by the National Optical Astronomy Observatory, which is
operated by the Association of Universities for Research in Astronomy
(AURA) under cooperative agreement with the National Science Foundation
(NSF).}} and STSDAS{\footnote{STSDAS: the Space Telescope Science Data 
Analysis System.}}.
 Cosmic rays were removed and the images in each filter were combined using 
the CRREJ routine. We found that all exposures in a given filter 
coregistered to better than $\sim$ 0.2 pixels.

\begin{figure}
\figurenum{1}
\epsscale{0.9}
\plotone{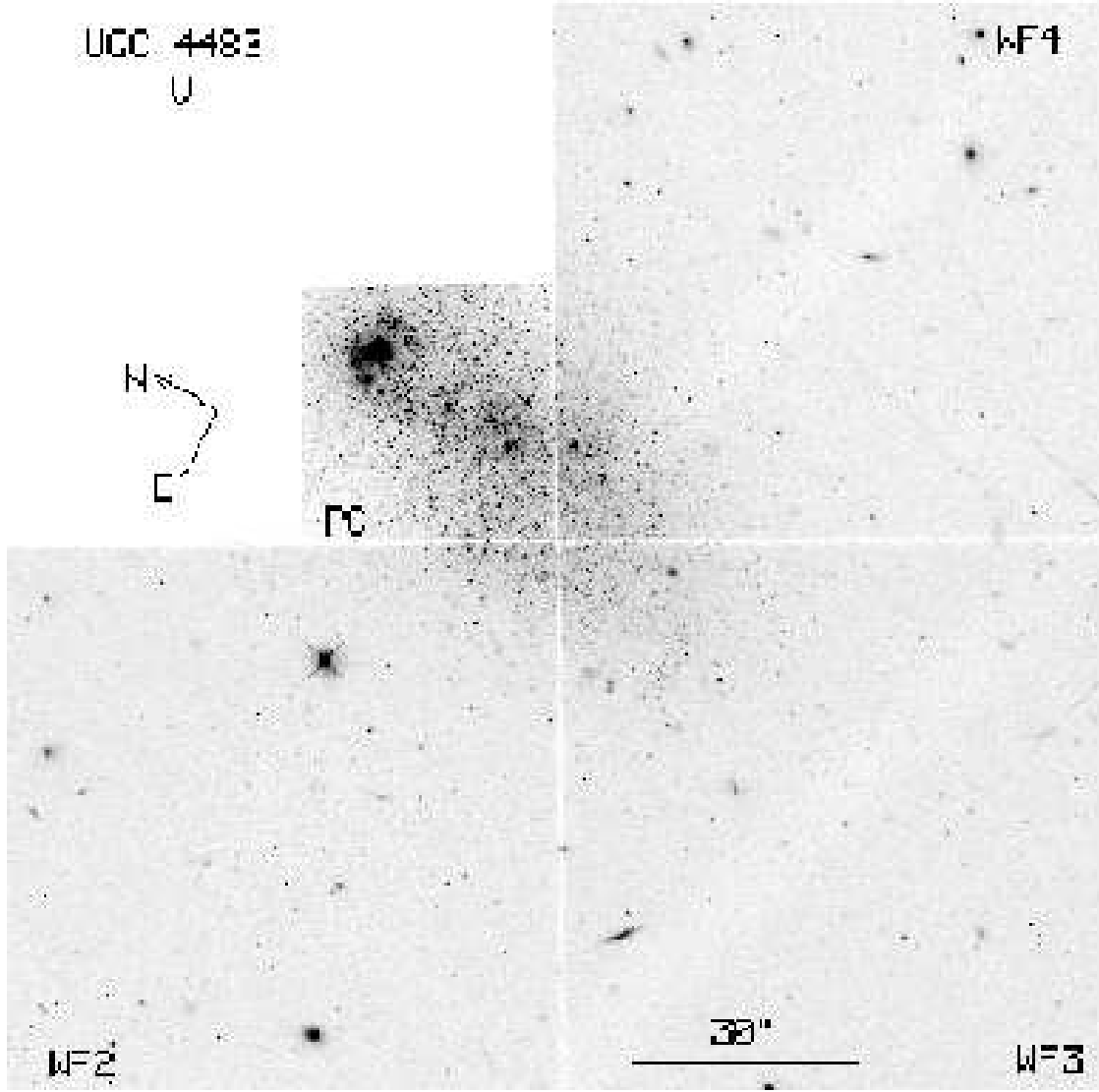}
\caption{{\sl HST} $V$ mosaic image of UGC 4483. 
The regions of the ongoing and recent star formation are at the northern
edge of the galaxy in the PC frame. Many background galaxies are also seen 
suggesting that UGC 4483 is projected onto a distant cluster of galaxies.
\label{Fig1}}
\end{figure}

\begin{figure}
\figurenum{2}
\epsscale{0.45}
\plotone{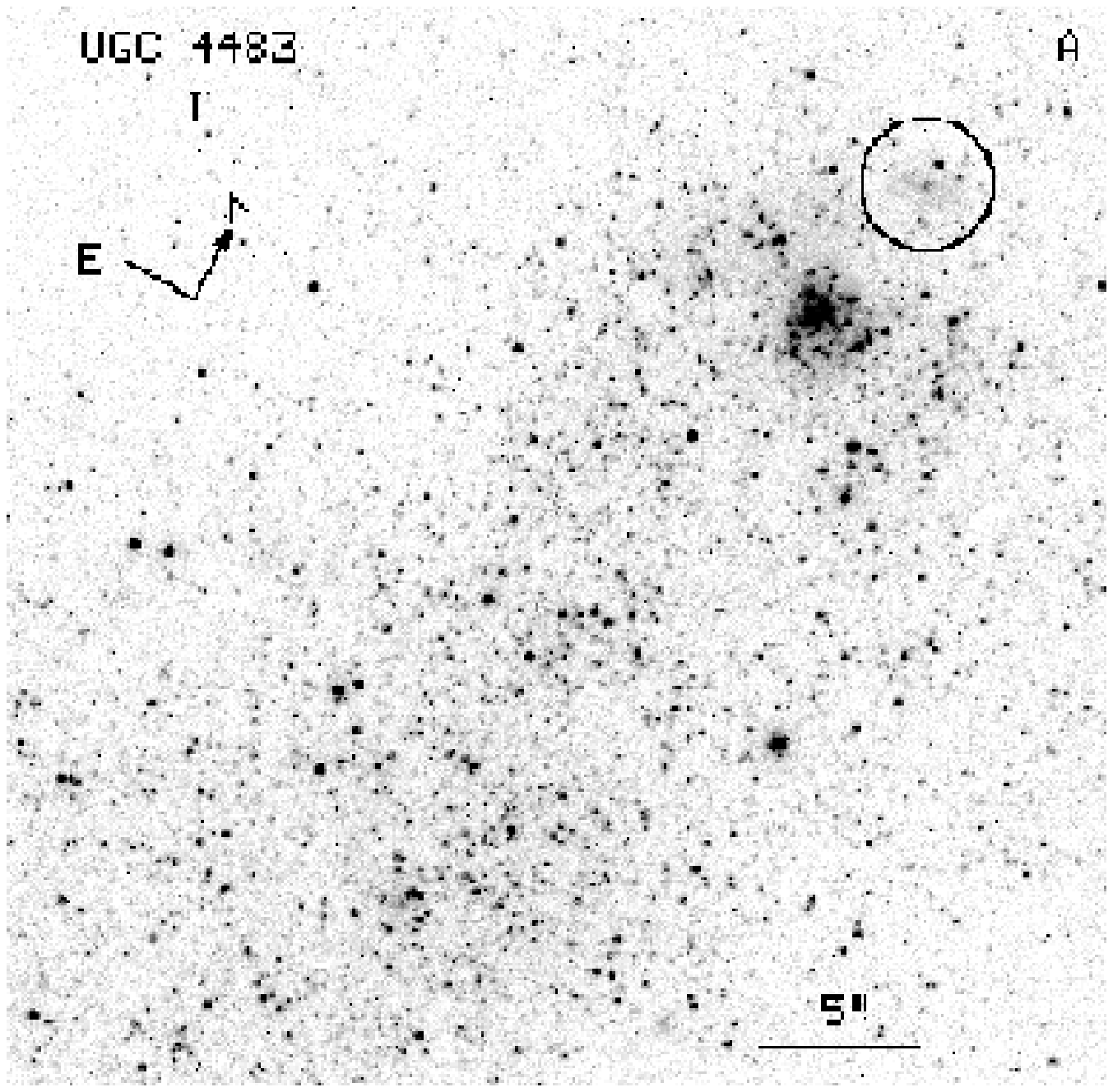}
\plotone{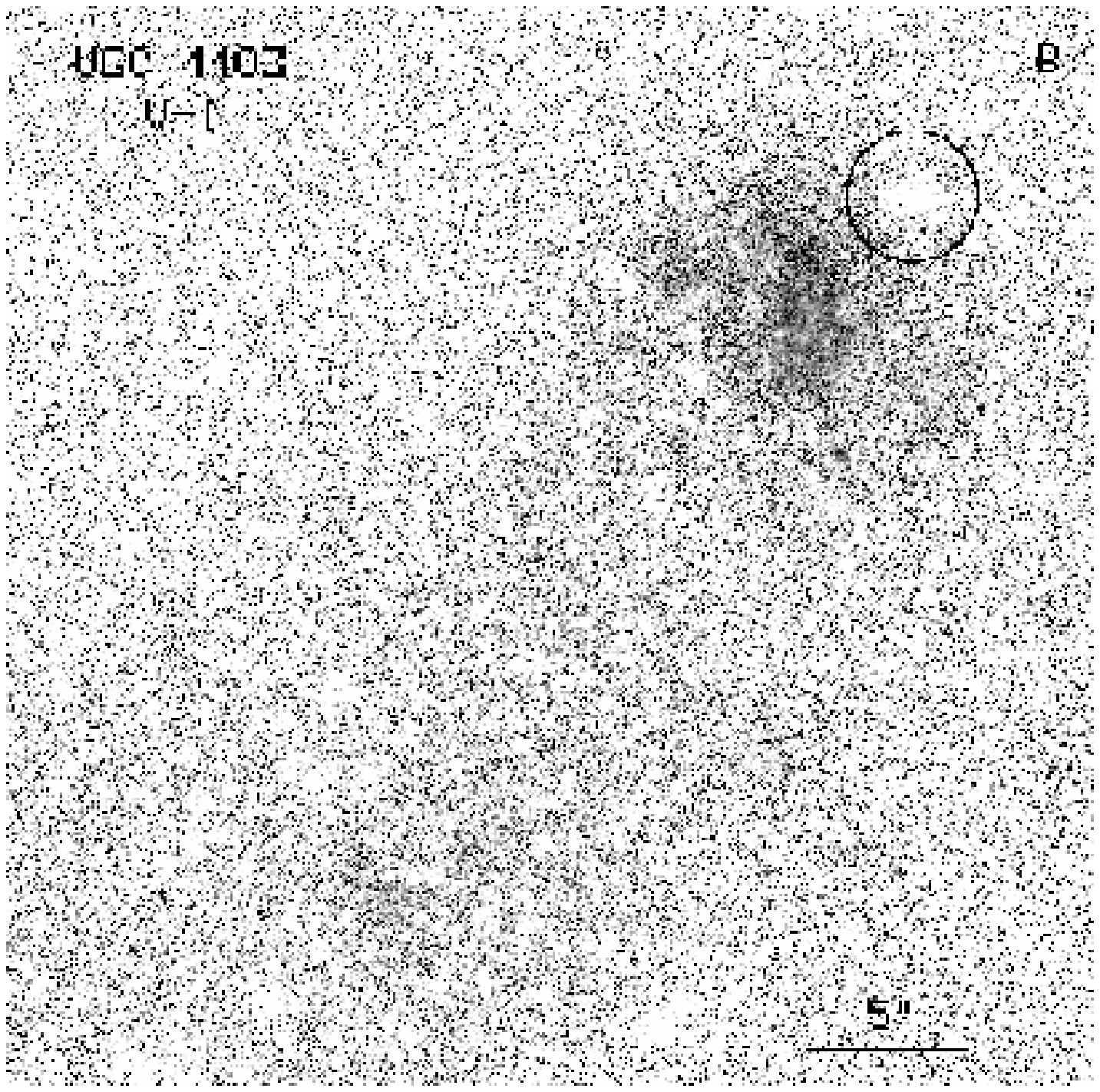}
\caption{PC $I$ (a) and $V-I$ (b) images.
In the $V-I$ image blue is black and red is white. Enclosed within the 
circle is a red regularly shaped background galaxy. The brightness 
distribution is clumpy implying that differential dust extinction is likely 
to be present. \label{Fig2}}
\end{figure}

In Fig. \ref{Fig1} is shown the mosaic $V$ image of UGC 4483. 
 Many faint background galaxies are visible in the field,
with several such galaxies along the line of sight to UGC 4483. The most 
prominent projected galaxy is located near the brightest star cluster. 
It is red, so is better seen in $I$ and $V-I$ images. In Fig. \ref{Fig2} we 
show PC $I$ and $V-I$ images where the prominent background galaxy is marked 
by a circle. The dark regions in the $V-I$ image are young regions with 
ongoing star formation and ionized gas emission. They have a patchy structure 
implying non-uniform dust extinction. 

   The superior spatial resolution of the {\sl HST} WFPC2 images combined 
with the proximity of UGC 4483 permits to resolve individual stars and study 
stellar 
populations in this galaxy by means of CMDs. We used the DAOPHOT 
package in IRAF for point-spread-function (PSF) fitting photometry. 
The PSFs are derived using the brightest stars in each frame.
There are many bright and relatively uncrowded stars in the PC frame, 
resulting in a reliable PSF. The PSFs for the WF frames
are not as reliable because of the lack of bright isolated stars.
The  measured full
widths at half maximum (FWHM) of the stellar profiles are
different for the PC and WF frames and for different filters. 
For the PC frame we obtain FWHMs
of 1.8 pixels and 2.3 pixels respectively for the $V$ and $I$ images.
The corresponding FWHMs for the WF frames are 1.5 pixels and 1.7 pixels. 
The background level in the PC field was measured 
in an annulus with radii 21 and 25 pixels (0\farcs97 and 
1\farcs15) around each source and 
subtracted. These radii are large enough 
so that the background annulus is free of contamination from the 
extended wings of the stellar profiles. We have checked that this is  
the case by using a background annulus closer to the source, with 
radii of 11 and 15 pixels (0\farcs51 and 
0\farcs69). The measured magnitudes with the larger annulus are
$\la$ 0.005 mag brighter than those obtained with the smaller annulus,
implying that the wings of the stellar profiles
do not contaminate the background appreciably at radii $>$ 0\farcs5.
The sky level in the WF fields was measured in 
an annulus with radii 11 and 15 pixels (1\farcs12 and 
1\farcs53). We also experimented with a
smaller background annulus with radii 6 and 11 pixels (0\farcs61 and 
1\farcs12) and did not find differences in the measured
magnitudes, implying again   
that the wings of the stellar profiles do not contaminate 
the background for radii $>$ 0\farcs5.
The zero point of the PSF-fitting 
stellar photometry was determined with the standard 
aperture photometry technique adopting a 2 pixel radius 
aperture in all frames. To convert instrumental magnitudes with an aperture
radius of 2 pixels to the corresponding magnitudes with 
the Holtzman et al. (1995a) calibrating aperture radius of
0\farcs5 (respectively 11 and 5 pixels for the PC and WF frames), 
we need to derive aperture corrections. For this, we compared 
PSF-fitted magnitudes of bright isolated 
stars with the magnitudes of the same stars
measured with the aperture photometry technique within an 0\farcs5 aperture.
For the PC frames we obtained the corrections $V_{\rm ap}$(0\farcs5) -- 
$V_{\rm fit}$ = --0.37 mag and $I_{\rm ap}$(0\farcs5) -- $I_{\rm fit}$ 
= --0.60 mag. For the WF frames the corresponding corrections are --0.22 mag
in $V$ and --0.23 mag in $I$. 
Our derived corrections are very similar to those obtained
by Holtzman et al. (1995a).

\begin{figure}
\figurenum{3}
\epsscale{0.8}
\plotone{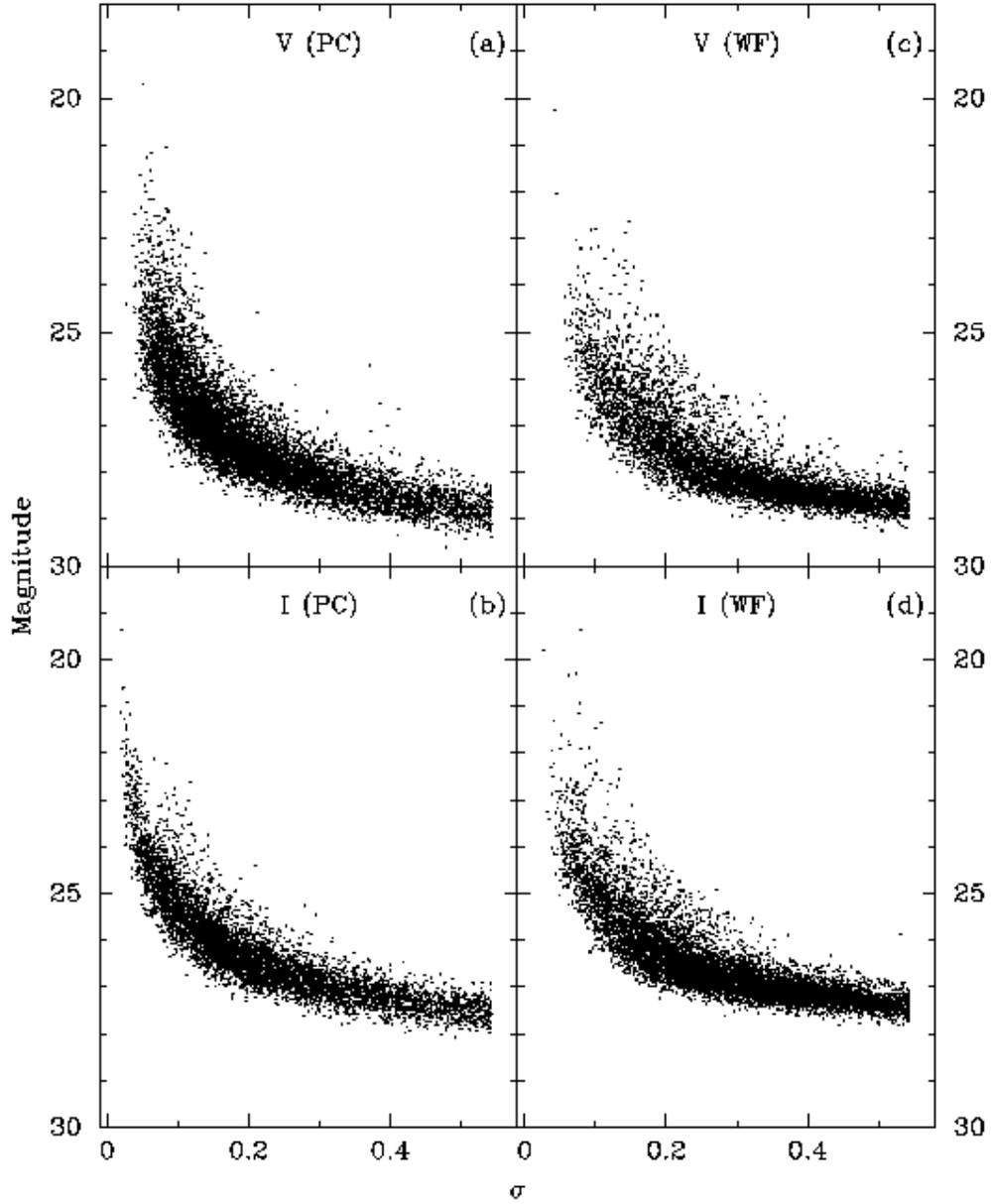}
\caption{Photometric error $\sigma$ as a function of apparent
magnitude in both $V$ and $I$ images for the PC frame (a and b)
and for the combined WF2,WF3,WF4 frames (c and d). The $V$ photometry is
$\sim$ 1 mag deeper than the $I$ photometry, going down to a limiting
magnitude of 29 mag. \label{Fig3}}
\end{figure}

Stars with $\chi^2$ greater than 4.0 and a sharpness out of the --1.0 -- +1.0
range in both PC and WF frames were eliminated
to minimize the number of false detections.
Correction for charge-transfer efficiency (CTE) loss has been
done according to the prescriptions of Dolphin (2000a).
Figure \ref{Fig3} shows the distribution of photometric errors as a 
function of $V$ and $I$ magnitudes as determined by DAOPHOT.
It is seen that errors are about 0.2 mag at $V$ = 27.5 mag and 
$I$ = 26.5 mag for both PC and WF frames. 
They increase to about 0.4 mag at $V$ = 28.5 mag and 
$I$ = 27.5 mag. 

We have excluded stars with photometric errors larger than 0.2 mag in the
final photometric list
for both $V$ and $I$ images. The total numbers of recovered stars in both PC 
and WF frames are respectively 7672, 5411 and 3853 in 
the $V$ image, the $I$ image and in both of these images at the same time. The 
corresponding numbers of recovered stars in the PC frame only are 4508, 2985
and 2250. We have adopted a matching radius of 1 pixel, although changing that 
radius within the range 0.5 - 2 pixels does not change the number of 
stars in both $V$ and $I$ frames appreciably. That nearly half the stars are 
not matched is likely to be caused by incompleteness effect and
an increasing number of false
detections at faint magnitudes. Indeed, the number of the stars with
$V$ = 25 -- 26 mag recovered in both $V$ and $I$ images is
$\sim$ 85\% that recovered in the $V$ image alone, while the number of
the stars with $V$ = 27 -- 28 mag recovered in both $V$ and $I$ images
is only $\sim$ 15\% that recovered in the $V$ image alone.

The transformation of instrumental magnitudes to the Johnson-Cousins $UBVRI$ 
photometric system as defined by Landolt (1992) was performed according to the 
prescriptions of Holtzman et al. (1995b) and Dolphin (2000a). The magnitudes 
and colors of
point sources were corrected for Galactic interstellar extinction adopting 
$A_V$ = 0.11 mag (Schlegel, Finkbeiner \& Davis 1998).

We have carried out a completeness analysis for the PC and WF4 frames
using the DAOPHOT routine 
ADDSTAR. For each frame we have added artificial stars amounting to $\sim$
5\% of the real stars detected in each magnitude bin in the original image.
We then performed a new photometric reduction 
using the same procedure as the one applied to the original frame, and checked 
how many added stars were recovered. This operation was repeated 10
times for each frame and each magnitude bin and the results were averaged. 
The completeness factor in each magnitude bin 
defined as the percentage of recovered artificial stars is shown 
in Table 1.
It can be seen that the completeness limit is reasonably good: it is about 
67\% -- 70\% in $V$ and 45\% -- 55\% in $I$ in the 26 -- 27 mag range,
but drops to $\sim$ 41\% -- 44\% and $\sim$ 7\% -- 15\% in the 27 -- 28 mag 
range.

\section{Distance determination}

\begin{figure}
\figurenum{4}
\epsscale{0.8}
\plotone{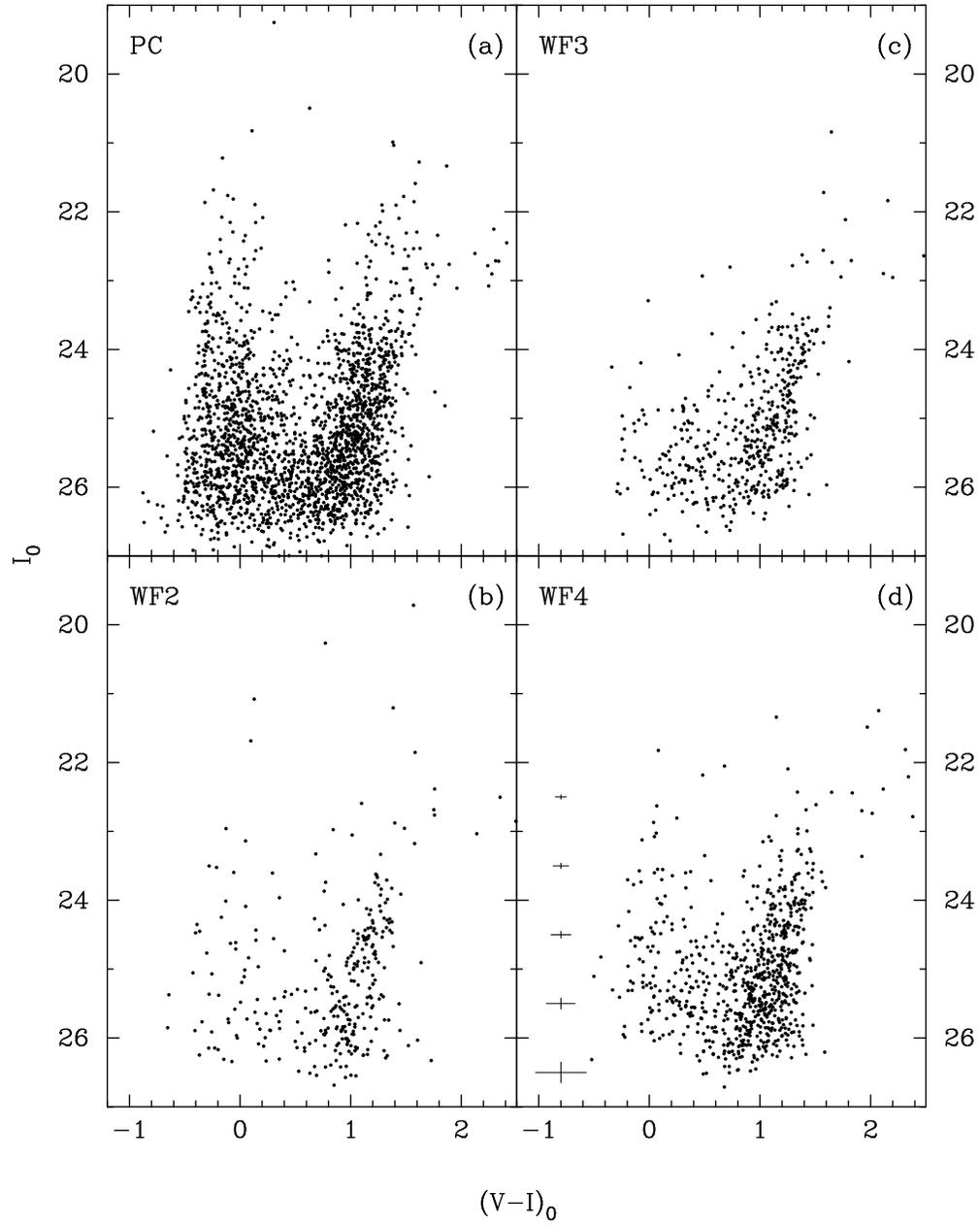}
\caption{$(V-I)_0$ vs $I_0$ color-magnitude diagrams in each
of the PC and WF frames. The observed magnitudes and colors are transformed
to the standard Johnson-Cousin $VI$ photometric system and corrected
for aperture and extinction adopting $E(B-V)$ = 0.11 mag. 
The photometric errors are shown by vertical and horizontal bars in panel (d).
\label{Fig4}}
\end{figure}

   In Fig. \ref{Fig4} we show $I_0$ vs $(V-I)_0$ CMDs
of UGC 4483 in all WFPC2 frames. The CMD for the PC frame is well populated by
both young main sequence stars and older red giant branch (RGB) and asymptotic
giant branch (AGB) stars. Such populations
have also been detected and discussed by Dolphin et al. (2001)
from less deep $V$ and $I$ {\sl HST}/WFPC2 images. The CMDs in the WF frames
are less populated, with a dominant contribution from RGB and AGB stars,
although younger stellar populations are present. 

The detection of RGB
stars allows in principle to derive the distance to the galaxy using
the apparent magnitude $I_0$ of the tip of RGB stars (TRGB). This
technique is based on the observed constancy of the
absolute magnitude $M_{I0}$ $\approx$ --4.0 mag of TRGB stars
in old globular stellar clusters 
(Da Costa \& Armandroff 1990). There is evidently a
sharp drop in the RGB distribution toward brighter $I_0$ magnitudes in the PC,
WF2 and WF3 frames (Fig. \ref{Fig4}a -- \ref{Fig4}c). However the drop is not 
so evident in the WF4 frame (Fig. \ref{Fig4}d) because of the large number 
of AGB stars atop the RGB tip. Therefore, care has been taken to minimize the 
contribution of AGB stars and more massive red helium burning (RHeB) stars
in the determination of $I_0$(TRGB). For all frames,
we have derived the distribution of the RGB stars located in the CMD to the 
red of the straight line defined by the pair of points $I_0$,$(V-I)_0$ = 
(27,0.4) and (22,1.5) to exclude the contribution of the RHeB stars, and to 
the blue of the straight line defined by $I_0$,$(V-I)_0$ = (27,0.8) and 
(22,1.9) to reduce the contribution of the AGB stars. 

\begin{figure}
\figurenum{5}
\epsscale{0.8}
\plotone{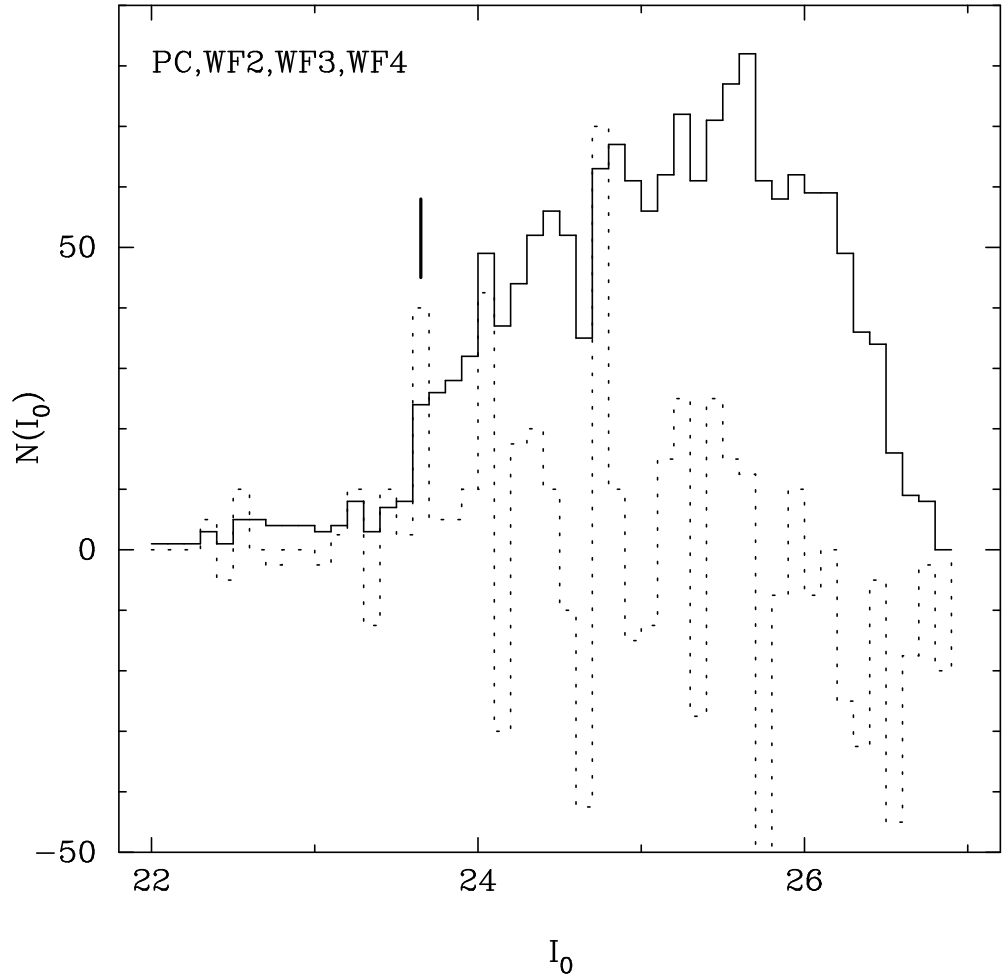}
\caption{Distribution of red giant branch stars as a function
of apparent $I_0$ magnitude (solid line) in the combined PC, WF2, WF3, WF4 
frames. The dotted line shows the derivative of the 
RGB star number distribution. The location of the RGB tip is marked by
vertical line and corresponds to $I_0$ = 23.65 mag. \label{Fig5}}
\end{figure}

The distribution of stars in the combined PC, WF2, WF3, WF4 frames in steps
$\Delta$$I_0$ = 0.1 mag is shown in Fig. \ref{Fig5} by a solid 
line. The derivative of the distribution is shown by a dotted line. 
The location of the TRGB is determined by the first large increase in both 
the RGB luminosity function and its derivative, 
and is shown by a vertical tick 
mark. The $I_0$(TRGB) 
is 23.65 $\pm$ 0.10 mag. We also experimented with luminosity functions
produced with different binnings, 
$\Delta$$I_0$ = 0.2 and 0.4 mag. The larger $\Delta$$I_0$
result in more monotonic luminosity functions, but the value of
the $I_0$(TRGB) is unchanged. It  
is 0.09 mag fainter, but is consistent within the errors 
with the one of 23.56 $\pm$ 0.10 mag derived by Dolphin et al. (2001)
from their short-exposure WF observations.

The distance modulus is given by 
$(m-M)_I$ = $I_0$(TRGB) -- $M_{I0}$(TRGB) with the absolute $I$ magnitude
defined as $M_{I0}$(TRGB) = $M_{\rm bol}$(TRGB) -- $BC$($I$), where 
$M_{\rm bol}$(TRGB) is the bolometric magnitude of the TRGB and $BC$($I$)
is the bolometric correction to the $I$ magnitude.
$BC$($I$) is related to the
$(V-I)$ color of the TRGB by $BC$($I$) = 0.881 -- 0.243$(V-I)_{\rm TRGB}$,
while the bolometric magnitude is related to the stellar metallicity
[Fe/H] by $M_{\rm bol}$(TRGB) = --0.19[Fe/H] -- 3.81
(Da Costa \& Armandroff 1990; Lee, Freedman \& Madore 1993). 
Usually [Fe/H] is derived 
using the $(V-I)$ color of RGB stars with absolute magnitude $M_{I0}$ = --3.
As pointed out by Lee et al. (1993) from analysis
of Yale theoretical isochrones, this is because $(V-I)$ is mainly
dependent on metallicity and not very much on age. We caution however
that this method for estimating metallicities of RGB stars may be less 
reliable when applied to relatively young stellar populations like those in
UGC 4483 (see section 4.2), because Lee et al. (1993) have restricted their 
analysis to stellar ages larger than 7 Gyr.
Futhermore, the Yale isochrones do not give a good fit to
the observed isochrones for globular clusters (see the discussion by 
Da Costa \& Armandroff 1990). 

Fig. \ref{Fig6}a shows the RGB region as
obtained from the PC frame. We adopt a distance modulus $m-M$ = 27.63,
as discussed later. The superimposed solid lines are, in order of increasing 
metallicity from left to right, the observed isochrones of 
the Galactic globular clusters M15, NGC 6397, M2, NGC 6752, NGC 1851 and 
47 Tuc with respective metallicities [Fe/H] = --2.17, 
--1.91, --1.58, --1.54, --1.29 and --0.71 (Da Costa \& Armandroff 1990).
Fig. \ref{Fig6}b shows the RGB region of the combined CMD for the 
WF2, WF3 and WF4 frames. 

   It is seen from Fig. \ref{Fig6}a that the $(V-I)$ colors of the RGB stars
in UGC 4483 are $\sim$ 0.1 mag bluer as compared to those in   
the most metal-deficient globular cluster M15 ([Fe/H] = --2.17)
in the Da Costa \& Armandroff (1990) sample. For the TRGB, we 
derive $(V-I)$ $\approx$ 1.3 mag, or $\sim$ 0.1 mag bluer than in the 
Local Group irregular galaxy Leo A, suggested by Tolstoy et al. (1998) to
contain a predominantly young stellar population. The $(V-I)$ color of the RGB 
stars at $M_{I0}$ = --3 mag is $\sim$ 1.1 mag. These blue colors 
may result from both metallicity and age effects. Assuming that
metallicity is the dominant effect we derive [Fe/H] = --2.39, using the 
equation [Fe/H] = --15.16 + 17.0$(V-I)_{-3}$ -- 4.9$(V-I)^2_{-3}$, where 
$(V-I)_{-3}$ is the color of RGB stars with $M_{I0}$ = --3 mag (Da Costa 
\& Armandroff 1990). The color-derived metallicity of the RGB stars in
UGC 4483 is $\sim$ 1.0 dex lower than the one derived in the H {\sc ii} region.
We caution however that this relation has been obtained for metallicities
in the range --2.1 $<$ [Fe/H] $<$ --0.7, so that extrapolation to
lower metallicities might be uncertain. 
For the metallicities --2.39 and --1.4 
(the metallicity of the ionized gas), $M_{I0}$(TRGB) = --3.93 and 
--4.03 mag respectively. Then the
respective distance moduli are $(m-M)$ = 27.58 and 27.68. Both  
values are slightly larger, but not significantly,
 than the value of 27.53 derived by Dolphin et al. (2001).
We adopt $(m-M)$ = 27.63 $\pm$ 0.12, as the mean of the two determinations. 
The distance to UGC 4483 is then $D$ = 3.4 $\pm$ 0.2 Mpc, which agrees within 
the errors  
with the distance of 3.2 $\pm$ 0.2 Mpc obtained by Dolphin et al. (2001).

   The above distance determination is based on the assumption that the RGB 
stars in UGC 4483 are $\sim$ 10 Gyr old. If they are younger, the
above method breaks down, and the metallicity of the RGB stars can be
considerably larger. Then, theoretical isochrones should be considered to
determine ages. Padua and Geneva evolutionary tracks are most often used for 
the analysis of resolved stellar populations in Local Group and more distant 
galaxies. However, as discussed e.g., Da Costa \& Armandroff (1990) and 
Lynds et al. (1998) the theoretical isochrones do not fit the observed ones
for globular clusters. We thus need to check the applicability of the
theoretical isochrones for the analysis of the RGB population. For that 
purpose, we compare theoretical 10 Gyr isochrones as calculated by the Padua
and Geneva groups for different metallicities with those observed for 
galactic globular clusters. 

   In Fig. \ref{Fig7} we show theoretical isochrones for two values of 
the heavy element mass fraction $Z$ = 0.0004 and $Z$ = 0.001 
(Bertelli et al. 1994 (Padua); Lejeune \& Schaerer 2001 (Geneva)) (thin solid 
lines) with the observed isochrones from Da Costa \& Armandroff (1990) for 
the globular clusters M2 with $Z$ = 0.0004 and NGC 1851 with $Z$ = 0.001 
(thick solid lines). All isochrones are plotted using a distance modulus 
$(m-M)$ = 27.63 to fit the CMD of UGC 4483. 
It is seen from Fig. \ref{Fig7}a and \ref{Fig7}c that the 10 Gyr 
theoretical isochrones with $Z$ = 0.0004 from either group do not fit 
the observed ones. The
Padua 10 Gyr isochrone is significantly bluer than the M2 isochrone, leading to
 a large age overestimate. Thus, the Padua isochrones would 
give an age of 10 Gyr for the RGB stars in UGC 4483, and an age larger than 
the age of the universe for M2. By contrast, the Geneva 10 Gyr isochrone is 
redder than that of M2, resulting in an underestimate of the age. In this 
case, a good fit to the RGB stars in UGC 4483 is given by the 
$\sim$ 2 Gyr isochrone, and the age of M2 would be 3 billion years.
                                              
The situation is better for models with $Z$ = 0.001 (Fig. \ref{Fig7}b and
\ref{Fig7}d). Both Padua and Geneva 10 Gyr isochrones fit the 
observed isochrone
of NGC 1851 quite well, the fit being slightly better for the Padua isochrones.
However, we note that the isochrones for $Z$ = 0.001 of Bertelli et al. (1994) 
are based on evolutionary tracks calculated with an older set of opacities. 
All other isochrones with different $Z$ given by those authors are calculated
with new opacities. Girardi et al. (2000) have also computed isochrones for
$Z$ = 0.001 with the new opacities. The 10 Gyr isochrones from Girardi et al.
(2000) do not fit the NGC 1851 isochrone,
being significantly bluer and resulting in a large age overestimate.


   Evidently, the uncertainties in theoretical isochrones for ages $\ga$ 1 Gyr
are large, and additional work is needed. 
We decided to adopt $(m-M)$ = 27.63 $\pm$ 0.12 for UGC 4483, 
derived from $M_{I0}$(TRGB) in globular clusters, assuming that it does not 
change appreciably between 1 and 10 Gyr. Although the distance of 
3.4$\pm$ 0.2 Mpc is not derived entirely in a self-consistent manner,
because of the age-metallicity 
ambiguity, it is the best that can be done with current observations and
models.

\section{Stellar populations}

   Several generations of stellar populations are present in the CMD of 
UGC 4483 suggesting continuous star formation over an extended period of time.
Ongoing star formation is evidenced by the presence of the 
high-surface-brightness supergiant 
H {\sc ii} region at the northern edge of the galaxy. Past star formation
is implied by the older stellar populations distributed over its whole 
low-surface-brightness body. In the CMD resulting from the combination of both 
PC and WF data (Fig. \ref{Fig8}), 
we have delimited the regions occupied by the different stellar populations:
the main-sequence (MS) stars with age less than $\sim$ 15 Myr, 
the blue loop (BL) and the red supergiant (RSG) stars 
with age between $\sim$ 20 Myr and $\sim$ 200 Myr, and the older AGB and 
RGB stars with age $\ga$ 1 Gyr.

   The spatial distributions of these different stellar populations are shown 
in Fig. \ref{Fig9}. The main sequence stars (Fig. \ref{Fig9}a) are nearly all
located in the H {\sc ii} regions at the northern edge of the BCD and show 
a clumpy distribution. The BL/RSG stars are found in and between the H {\sc ii}
regions and their distribution is smoother and more extended than that of the
MS stars (Fig. \ref{Fig9}b). Finally, AGB and RGB stars occupy
the largest area of the galaxy (Fig. \ref{Fig9}c -- \ref{Fig9}d). 
Similar results with somewhat smoother and more extended spatial
distributions of populations with different ages have also been 
found by Dolphin et al. (2001). Such a general trend of more
extended spatial distributions for older stellar populations is also
found in other dwarf galaxies observed with
{\sl HST} (e.g., Lynds et al. 1998; Schulte-Ladbeck et al. 1998; Crone et al.
2000) and is likely a consequence of the diffusion and relaxation of stellar
ensembles. We discuss next in detail the different stellar populations in 
UGC 4483.

\subsection{Young stellar populations}

   We restrict our discussion in this section to the PC image because
most of the ongoing and recent star formation is concentrated there, 
as evidenced by the bright supergiant H {\sc ii} region (Fig. \ref{Fig1})
and its blue $(V-I)$ color (Fig. \ref{Fig2}b). Fig. \ref{Fig10} displays
a zoom of the PC $V$ and $I$ images showing the ionizing compact stellar
cluster in the supergiant H {\sc ii} region. The stellar cluster has a
diameter $<$ 1\arcsec\ or $<$ 16.5 pc for our derived distance of 3.4 Mpc. 
Dolphin et al. (2001) have constructed a CMD for this cluster and derived
an age of 15 Myr for it. This age can not be correct, because with such an age 
the expected equivalent width of the H$\beta$ emission line in the H {\sc ii} 
region is less than 10\AA\ (e.g., Schaerer \& Vacca 1998), instead of the
value of 169\AA\ measured by Izotov et al. (1994). The latter value gives an
age of only 4 Myr. A similar H$\beta$ equivalent width has been obtained
by Skillman et al. (1994).

   We measure the $V$ and $I$ magnitudes for this cluster using a square 
aperture of 27 pixels (or 1\farcs2) on a side centered at the point with 
coordinates $x$ = 601 and $y$ = 581, where the origin $x$ = $y$ = 0 is at
the left lower corner of the PC chip. The galaxy background, measured at 
several points $\sim$ 3\arcsec\ away from the cluster and amounting to 350 
e$^-$ pixel$^{-1}$ in $V$ and 140 e$^-$ pixel$^{-1}$ in $I$, has been 
subtracted. We derive $V$ = 19.38 $\pm$ 0.01 mag, $I$ =
19.61 $\pm$ 0.01 mag, and $(V-I)$ = --0.23 $\pm$ 0.01 mag for the 
cluster. Such a blue
color is typical for young ionizing stellar clusters. We have not corrected 
the magnitudes and colors for the contamination of the ionized gas because 
a precise knowledge from spectroscopic observations of the emission line 
intensities in exactly the same region is required. The apparent $I$ 
magnitude of the cluster translates into an absolute magnitude of 
$\sim$ --8.0 mag, equivalent to the emission of $\sim$ 15 O stars with 
absolute magnitude $M_I$ = --5 mag. Approximately such a number of stars 
can be seen in the cluster (Fig. \ref{Fig10}). 

   A very blue compact source with coordinates $x$ = 625 and $y$ = 491 
is seen at a distance $\sim$ 4\arcsec\ from the cluster (within the circle
in Fig. \ref{Fig10}). It is embedded within the ionized gas envelope 
surrounding the cluster. This source was not recovered by DAOPHOT because of 
its large FWHM of 4.2 pixels, about two times the FWHM of point sources.
Using APPHOT and a 5-pixel aperture, we obtain $V$ = 21.83 $\pm$ 0.01 mag, 
$I$ = 22.38 $\pm$ 0.02 mag and $V-I$ = --0.55 $\pm$ 0.02 mag. With a smaller 
3-pixel aperture, the respective values are $V$ = 22.33 $\pm$ 0.02 mag, 
$I$ = 22.92 $\pm$ 0.03 mag and $V-I$ = --0.59 $\pm$ 0.04 mag. Adopting the 
latter values, the absolute $I$ magnitude of the source is --4.7 mag. 
Its $(V-I)$ color is bluer than the color of the most massive main-sequence
O stars ($\sim$ --0.3 mag). It is probably an early-type 
Wolf-Rayet star embedded in a H {\sc ii} region with a 4-pixel radius, or 
0\farcs18, corresponding to a linear radius of 3 pc. Several other
bright blue stars are seen around the cluster (compare Fig. \ref{Fig10}a
and \ref{Fig10}b). However, no appreciable ionized gas emission is present
in the vicinity of these stars, and they are probably 
late O or early B supergiant stars.

   To study the spatial variations of the stellar populations in the PC frame,
we have divided it into six regions as shown in Fig. \ref{Fig11}. 
Region 1 includes 
the supergiant H {\sc ii} region and contains many bright main-sequence
stars as is evident from examination of its CMD in Fig. \ref{Fig12}a.
In Fig. \ref{Fig12}, superimposed on the data are shown theoretical isochrones
for $Z$ = 0.001 (Bertelli et al. 1994). We use this isochrone set because its
metallicity is close to the one observed in the H {\sc ii} region, and its
10 Gyr isochrone fits well the observed isochrone of the globular cluster
NGC 1851 which has a similar metallicity (Fig. \ref{Fig7}b). 
However, as discussed in section 3, it must be kept in mind that 
the absolute age calibration of 
these isochrones is somewhat uncertain, and they are best used in a 
differential manner to compare stellar population ages in different regions.
The logarithms of
the age of each isochrone are given in Fig. \ref{Fig12}f.
 Older stars are present in region 1 as well. The presence of blue loop
and red supergiant stars indicates that star formation has continued in this
region for the last several 100 Myr. Finally, $\sim$ 2 Gyr red giant stars can
also be seen in the CMD of region 1. However, no AGB stars are seen.
We conclude that, in addition to a strong recent burst of star formation as 
evidenced by the presence of the supergiant H {\sc ii} region, star formation 
has occured during at least the last 2 Gyr in region 1. Alternatively, 
if the metallicity of the stars is significantly lower than that of the gas, 
then our age estimates could be significantly larger.

Similar conclusions regarding the star formation history can be made for 
regions 2 to 5 (Fig. \ref{Fig12}b -- \ref{Fig12}e), except that the main 
sequence is not as well populated as in region 1, and ionizing upper main 
sequence stars are absent. 

In summary, a large part of the galaxy around the supergiant H {\sc ii}
region shows evidence for continuous star formation during at least the 
last 2 Gyr. 
The exception is region 6 which is characterized by a low density of stars,
and where mainly $\ga$ 2 Gyr RGB stars are present. It is likely that star 
formation never occurred in this region and the RGB stars migrated there from
denser regions of the galaxy.

\subsection{The RGB and AGB stellar populations}

   RGB and AGB stars are good tracers of the star formation history of
UGC 4483 during the last several hundred Myr (AGB stars) to $\ga$ 2 Gyr 
(RGB stars). If there are no older stellar populations in the galaxy, then it
can be considered as a relatively 
young galaxy, since a 2 Gyr age is much lower than the 
age of the Universe. Unfortunately, 
the galaxy is too far away for {\sl HST} to detect the red giant clump and 
horizontal-branch stars which are good indicators of stellar populations
older than several Gyr.
Therefore, our discussion of the age of UGC 4483 will have to rely 
on the observed RGB and AGB stellar populations. It will be by necessity more 
qualitative than quantitative. This is because of the 
age-metallicity degeneracy of RGB star colors, 
the poor knowledge of the properties
of AGB stars and the uncertainties in the theoretical isochrones.

\begin{figure}
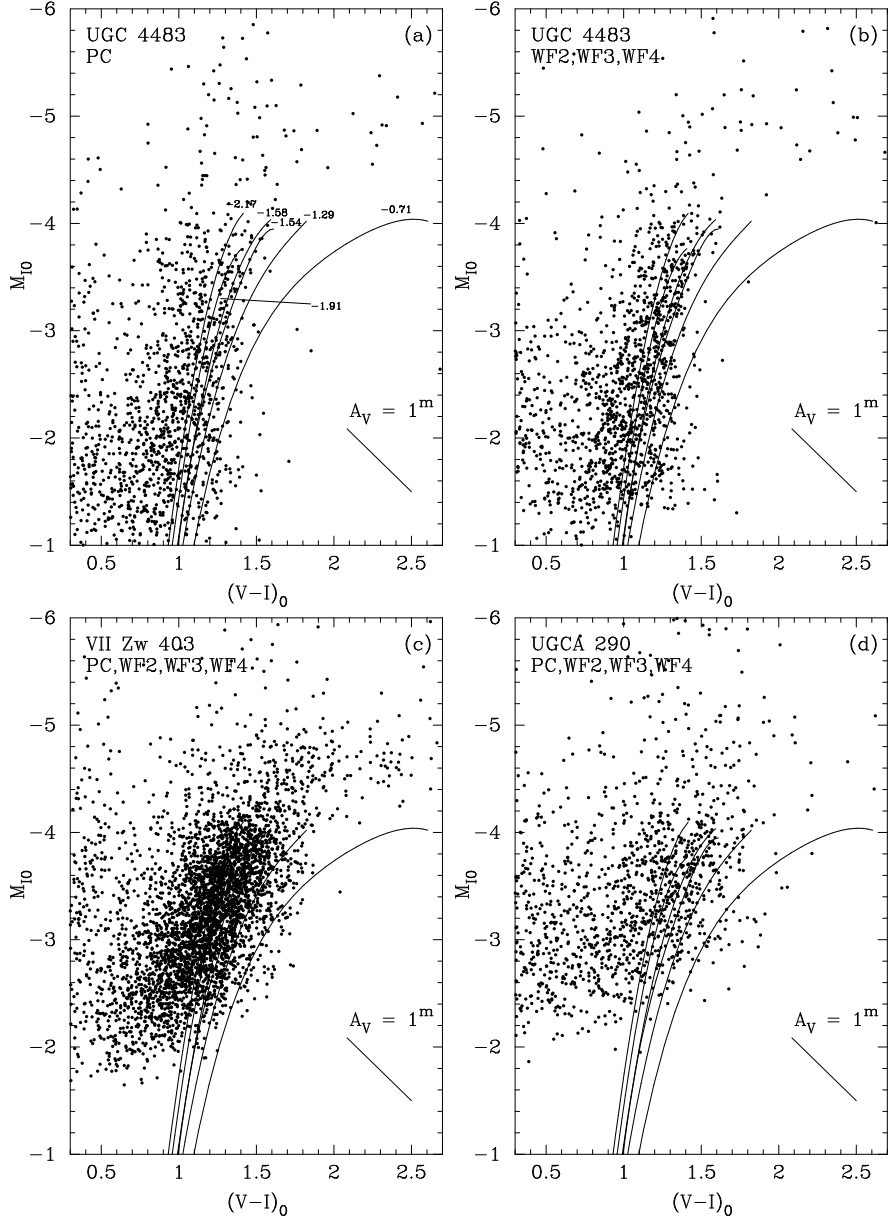

\figurenum{6}
\epsscale{0.35}
\plotone{Fig6a.ps}
\plotone{Fig6b.ps}
\plotone{Fig6c.ps}
\plotone{Fig6d.ps}
\caption{Absolute magnitude $M_{I0}$ vs $(V-I)_0$ diagram of
the RGB stars in UGC 4483 for 
(a) the PC frame and (b) the combined WF2, WF3, WF4 frames. 
The adopted distance modulus is $m-M$ = 27.63.
Magnitudes and colors are corrected for Galactic extinction with $A_V$ = 0.11
mag. 
In (c) and (d) are given similar plots
for the BCDs 
VII Zw 403 and UGCA 290. The 
distance modulus and extinction correction are taken from Lynds et al. (1998)
for VII Zw 403 and Crone et al. (2000) for UGCA 290.  
The superimposed solid lines are, from
left to right, the observed isochrones arranged in order of increasing 
metallicity of the globular clusters M15 
([Fe/H]=--2.17), NGC 6397 (--1.91), M2 (--1.58), NGC 6752 (--1.54), NGC 1851
(--1.29) and 47 Tuc (--0.71) (Da Costa \& Armandroff 1990). The metallicity 
labels are given in panel (a). 
\label{Fig6}}
\end{figure}

\begin{figure}
\figurenum{7}
\epsscale{0.45}
\plotone{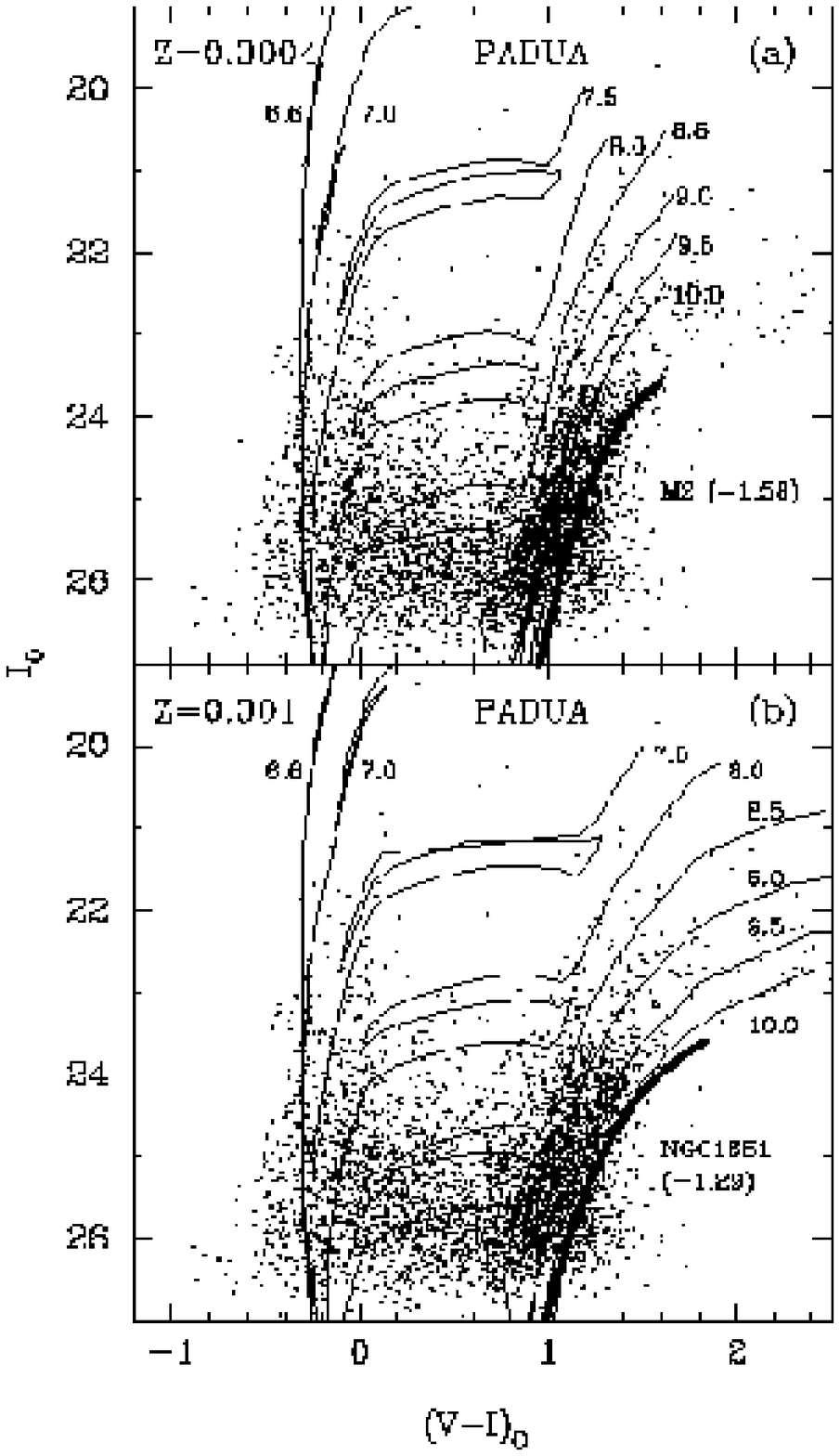}
\plotone{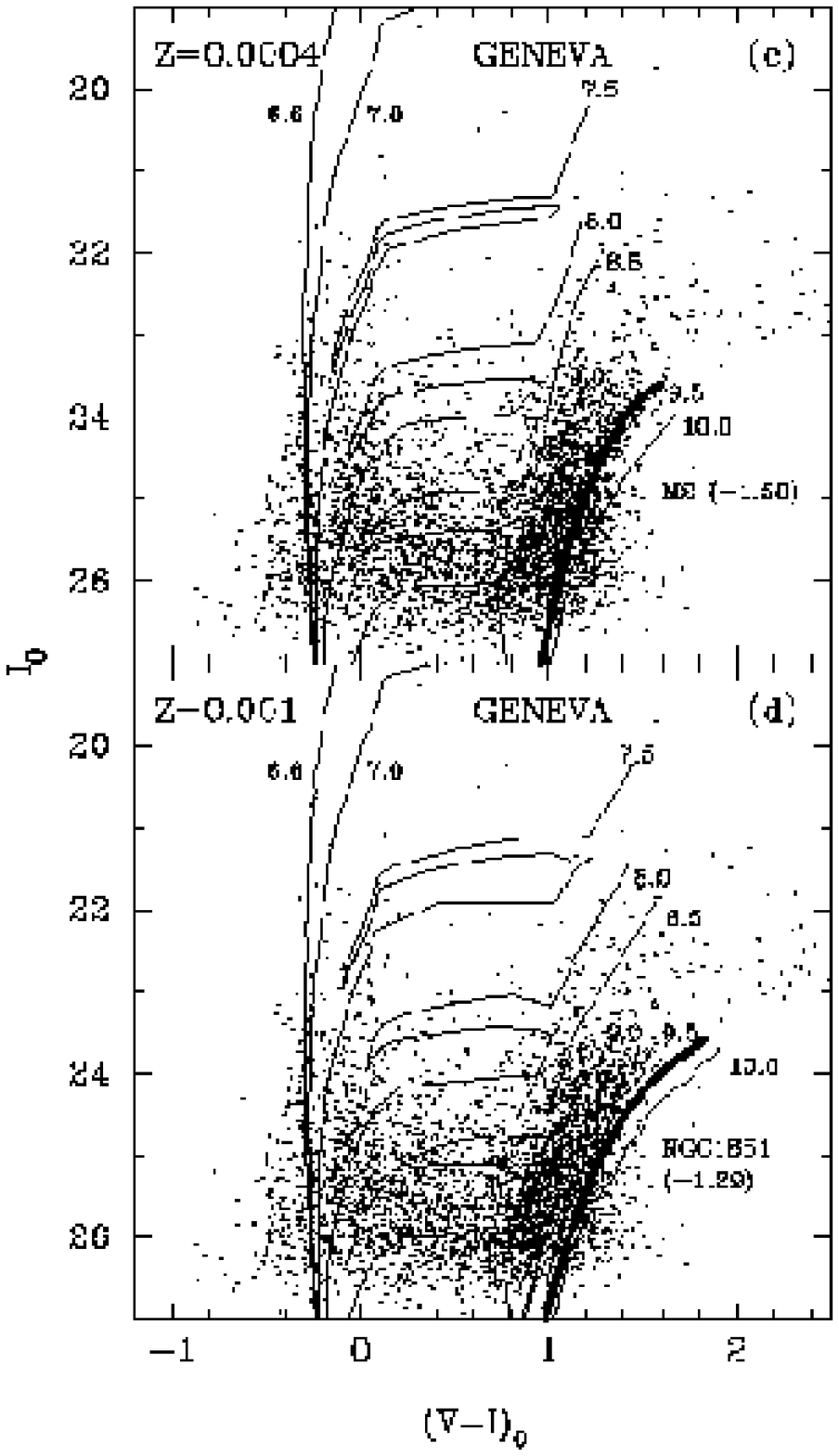}
\caption{Comparison of theoretical isochrones (thin lines) with observed ones 
of two globular clusters (M2 and NGC 1851) with matching metallicities (thick
lines) for two sets of models: Padua models with $Z$ = 0.0004 (a) and
$Z$ = 0.001 (b) (Bertelli et al. 1994), and Geneva models with 
$Z$ = 0.0004 (c) and $Z$ = 0.001 (d) (Lejeune \& Schaerer 2001). Each 
theoretical isochrone is labeled by the logarithm of the age in year.
The isochrones are overplotted on the CMD of UGC 4483. Only the Padua 
model with $Z$ = 0.001 fits reasonably well the isochrone for the 
globular cluster NGC 1851 (b). The Padua model with $Z$ = 0.0004 (a)
cannot fit the reddest part of the RGB in UGC 4483.
\label{Fig7}}
\end{figure}

\begin{figure}
\figurenum{8}
\epsscale{0.7}
\plotone{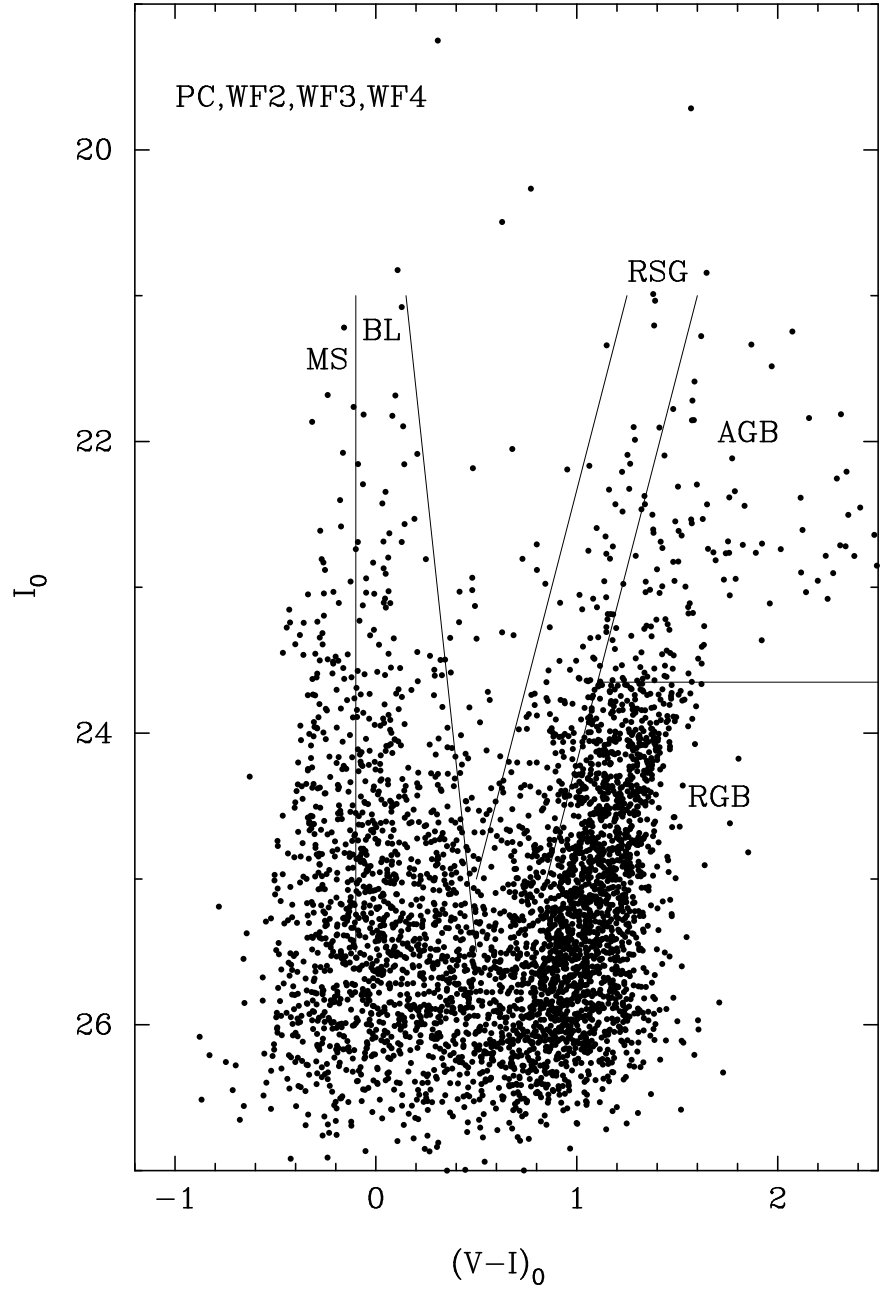}
\caption{Combined CMD of UGC 4483 overplotted by lines showing the
locations of main-sequence (MS), blue loop (BL), red supergiant (RSG),
red giant (RGB) and asymptotic giant branch (AGB) stars. 
\label{Fig8}}
\end{figure}

\begin{figure}
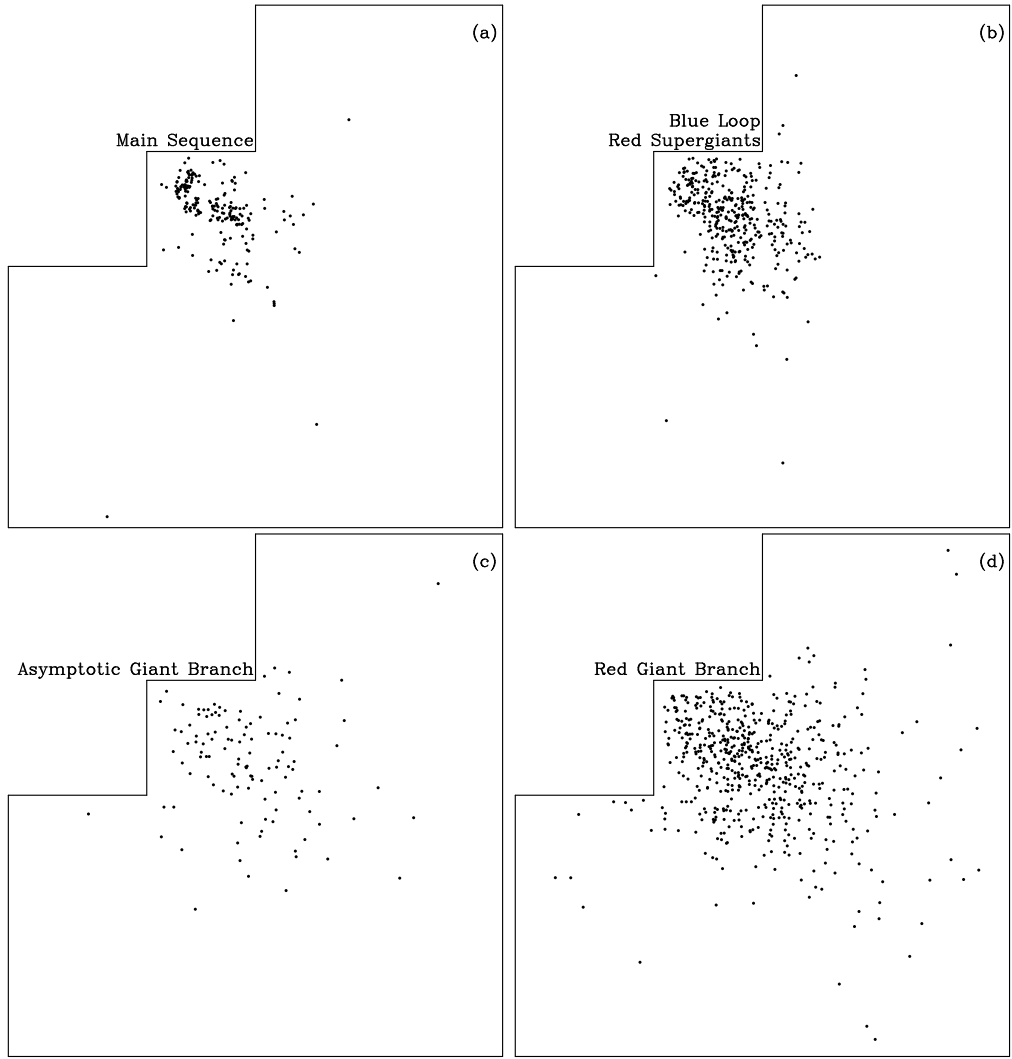

\figurenum{9}
\epsscale{0.4}
\plotone{Fig9a.ps}
\plotone{Fig9b.ps}
\plotone{Fig9c.ps}
\plotone{Fig9d.ps}
\caption{Spatial distribution of the (a) main-sequence, (b) blue 
loop and red supergiant, (c) asymptotic giant branch and (d) red giant branch
stars in UGC 4483. The main sequence stars show a clumpy distribution and are 
concentrated
mainly in the PC frame. Stars with progressively increasing ages [(b) to (d)] 
are distributed more smoothly and occupy successively larger volumes in the 
galaxy. The WFPC2 camera covers 160\arcsec\ on a side.
\label{Fig9}}
\end{figure}

\begin{figure}
\figurenum{10}
\epsscale{1.1}
\plottwo{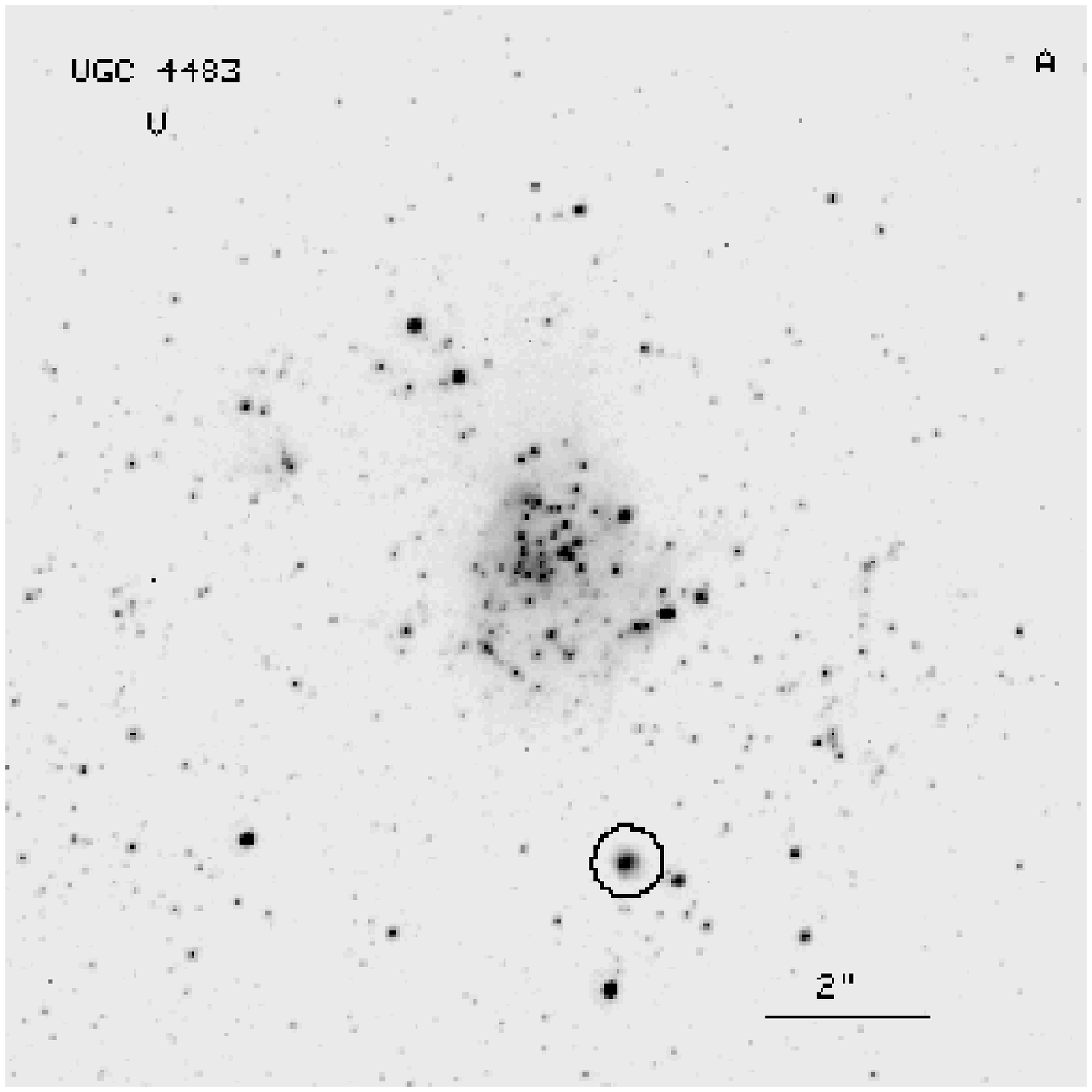}{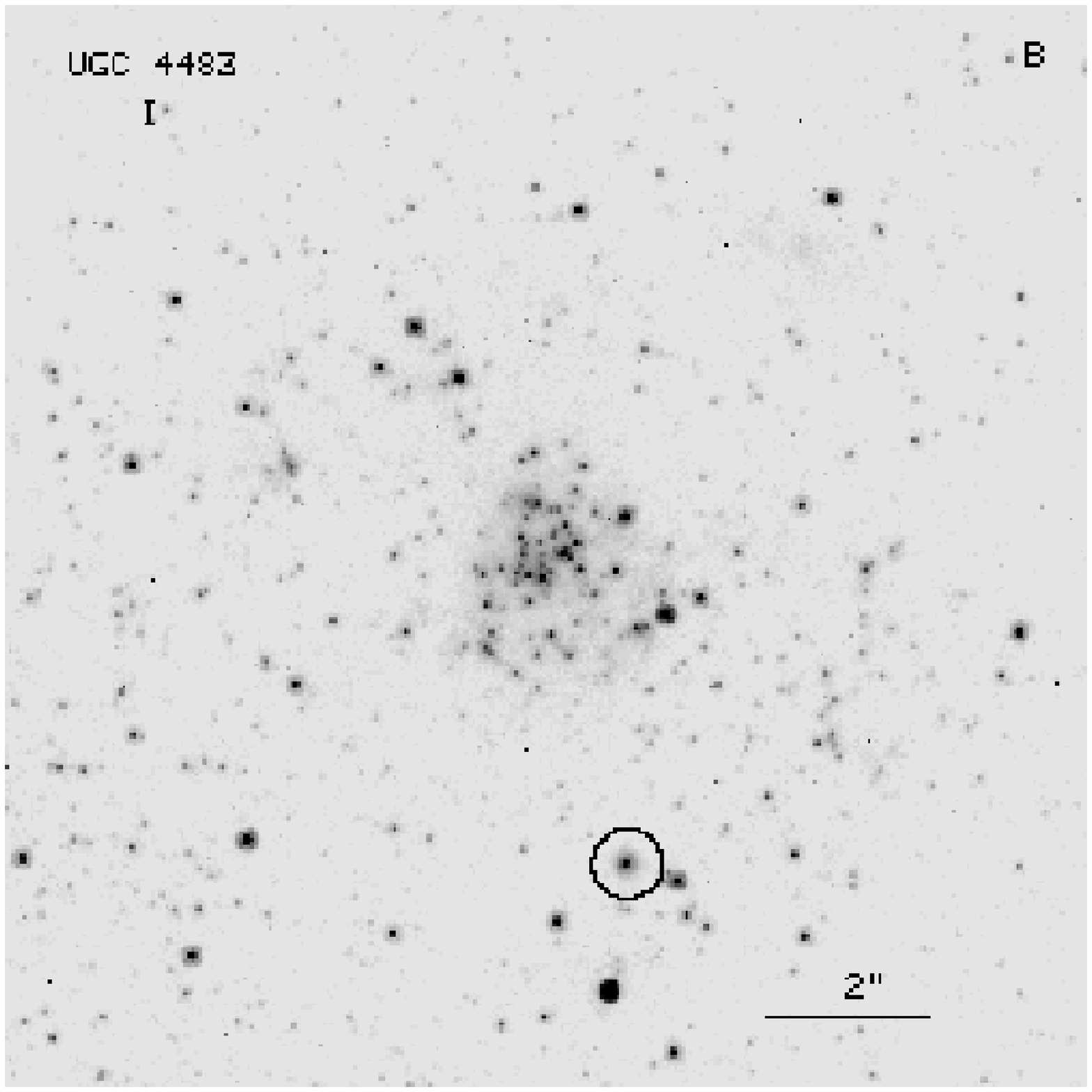}
\caption{Area of the PC image in $V$ (a) and $I$ (b) which contains
the brightest and youngest stellar cluster. Ionized gas is visible around
the cluster. The compact blue object marked by the circle is an H {\sc ii} 
region associated with a single hot, presumably of the Wolf-Rayet type. 
The orientation is the same as in Figure \ref{Fig2}. \label{Fig10}}
\end{figure}

\begin{figure}
\figurenum{11}
\epsscale{1.0}
\plotone{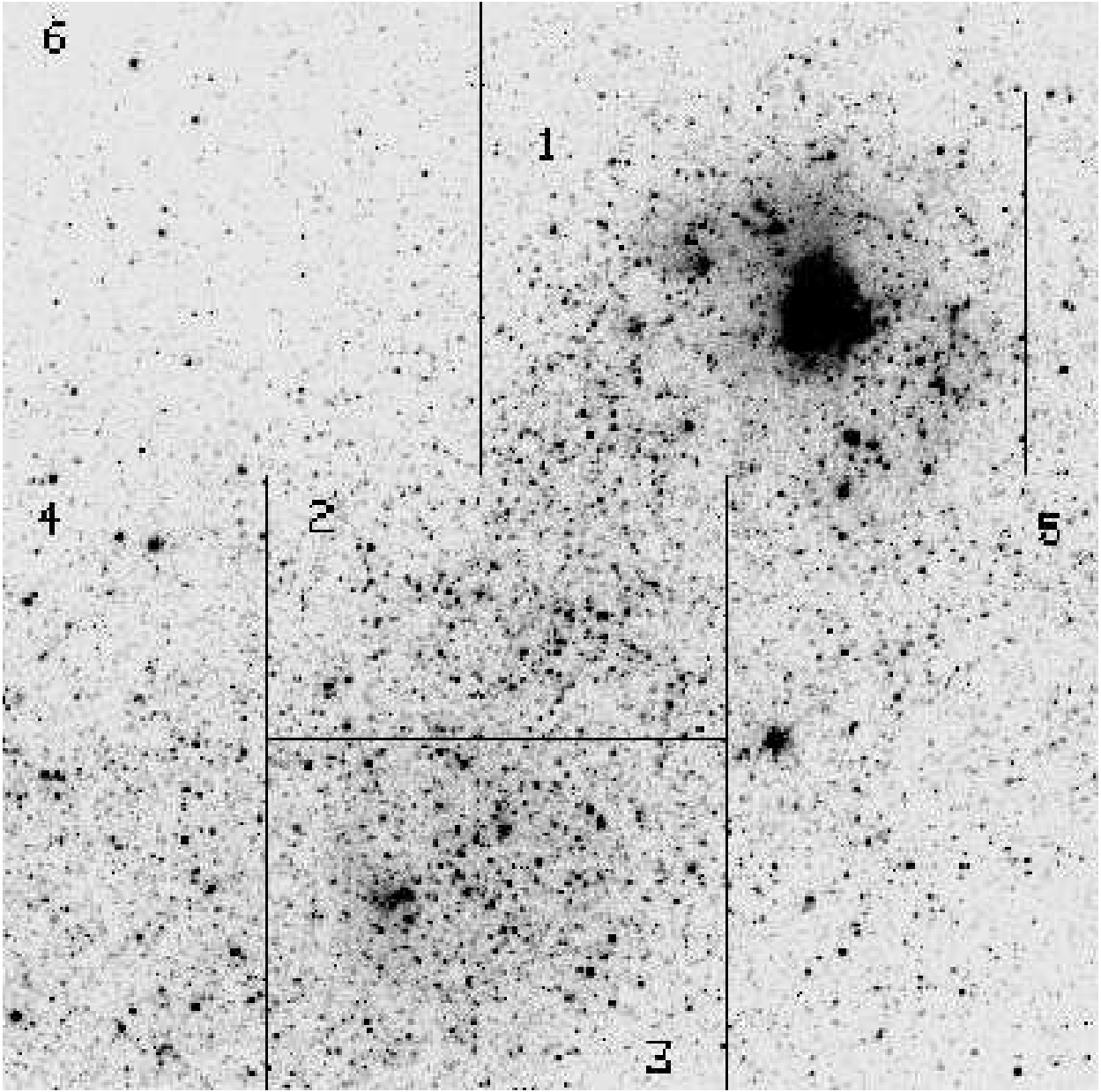}
\caption{PC $V$ image of UGC 4483 divided into 6 different regions for CMD 
analysis of the stellar populations (see Fig. \ref{Fig12}). The image is
36\farcs8 on a side. The orientation is the same as in Figure \ref{Fig2}.
\label{Fig11}}
\end{figure}

\begin{figure}
\figurenum{12}
\epsscale{1.0}
\plotone{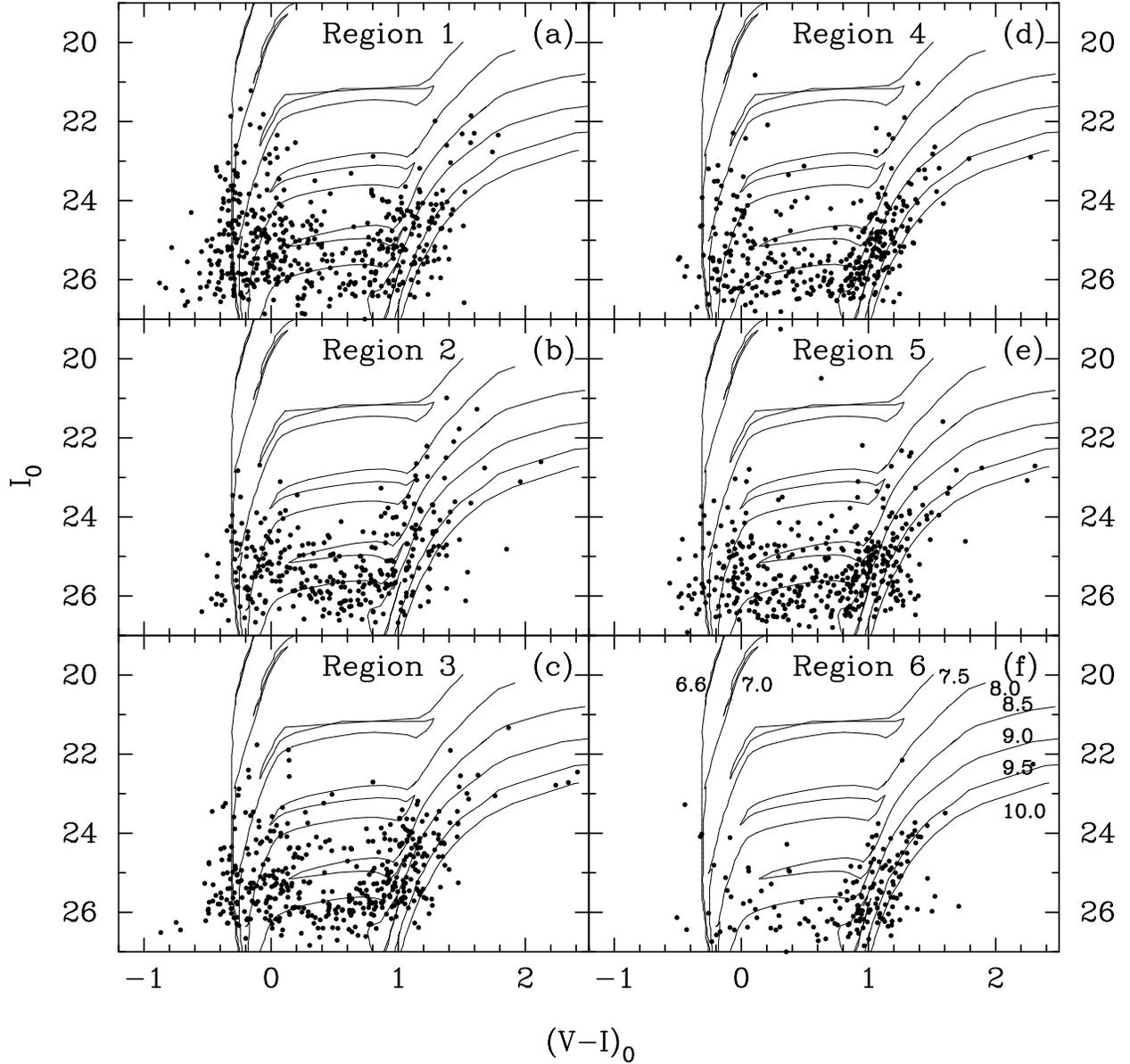}
\caption{$(V-I)_0$ vs $I_0$ CMDs for the 6 regions of UGC 4483 
shown in Fig. \ref{Fig11}. 
Superimposed are theoretical isochrones for a heavy element mass fraction 
$Z$ = 0.001 (Bertelli et al. 1994). The logarithms of the ages in years 
for each
isochrone are shown in (f). Populations of different ages are present in 
regions 1 -- 5 suggesting continuous star formation during the last 
$\sim$ 2 Gyr. Region 6 is populated only by a relatively
old stellar population of 300 Myr -- 2 Gyr.
\label{Fig12}}
\end{figure}

   We have already pointed out that the most striking feature in the CMD of 
UGC 4483 is that the RGB stars are very blue. The $(V-I)$ color of the TRGB 
as measured in the PC frame is only 1.3 mag (Fig. \ref{Fig6}a), bluer 
than the color of the TRGB of old Galactic globular clusters ($\ga$ 1.4 mag). 
The $(V-I)$ color of the TRGB in the WF frames, 
though slightly redder, is 1.35 mag, or 0.05 mag bluer than the TRGB color
of the globular cluster M15 with metallicity [Fe/H] = --2.17. Such blue
colors of the RGB stars can be explained in two ways: 1) by a very low 
metallicity [Fe/H] $\sim$ --2.4 and an old age, or 2) by a higher metallicity 
and a younger age ($<$ 10 Gyr). 

   Let's first consider case 1. If we suppose that the RGB stars in UGC 4483 
are 10 Gyr old, then the extremely low metallicity required to produce their
relatively blue colors is not easily understood. While stars with [Fe/H] 
$\sim$ --2.4 cannot be excluded in principle, they are rare in the 
Milky Way and in other well-studied nearby galaxies. There is an additional
problem: if the slightly redder colors of the RGB stars seen in the WF frames 
is a real effect due to a higher metallicity, why would stars seen in the
periphery of UGC 4483 which are likely to be older be more metal-rich 
as compared 
to the younger RGB stars in the main body of the galaxy? It is more plausible
that case 2 prevails, that the blue colors of the RGB stars in UGC 4483
are caused by their younger age rather than by their lower metallicity. 
In that case, the metallicity of the ionized gas, while being probably
higher than the stellar metallicity, may still be a reasonable approximation
for the metallicity of the RGB stars. Then, using models with $Z$ = 0.001
from Bertelli et al. (1994), we derive an age for the RGB stars 
of $\sim$ 2 Gyr. 
Uncertainties in the theoretical isochrones and the unknown metallicity of the
RGB stars in UGC 4483 preclude a more precise determination. While our CMDs 
do not show evidence for stellar populations older than $\sim$ 2 Gyr, they do 
not go deep enough to definitively rule them out.

   AGB stars are tracers of the intermediate and old stellar populations. 
They are not as numerous in UGC 4483 as, for example, in the BCD VII Zw 403 
(see Fig. 9 in Lynds et al. 1998) which has a slightly higher ionized gas 
metallicity ($Z$ $\sim$ $Z_\odot$/15). AGB stars
in UGC 4483 have a mean absolute $I$ magnitude $\sim$ --4.9 mag
(Fig. \ref{Fig6}a and \ref{Fig6}b) and are brighter by $\sim$ 0.3 mag than 
those in VII Zw 403 and UGCA 290 (Fig. \ref{Fig6}c and \ref{Fig6}d). Again, 
as in the case of RGB stars, the difference in the absolute brightness of the 
AGB stars in UGC 4483 can be due to two reasons:
a lower metallicity and/or a younger age than in VII Zw 403 and UGCA 290. 

There are no theoretical isochrones that can reproduce well the observed 
properties of AGB stars at low metallicities (e.g., Lynds et al. 1998).
While AGB stars in the CMDs of UGC 4483, VII Zw 403 and some Local Group 
irregular galaxies have a constant absolute magnitude $M_I$, the models 
predict a steady increase of $M_I$ with increasing $(V-I)$ color, to the tip 
of the AGB (TAGB) phase, when the brightness
of AGB stars reaches their maximum. The Padua theoretical isochrones with 
$Z$ $<$ 0.001 are too short and too steep to be able to reproduce the red 
colors of AGB stars (e.g., Fig. \ref{Fig7}a -- \ref{Fig7}b). 
As for Geneva isochrones, they do not include the AGB stage.
Therefore, instead of performing an absolute comparison of the theoretical
isochrones with the data points we can only perform a qualitative
differential comparison of the AGB populations seen in UGC 4483
and VII Zw 403, using the Bertelli et al. (1994) $Z$ = 0.001 models. 
Assuming that the AGB stars in both galaxies have the same metallicity,
the difference of 0.3 mag in the TAGB brightness translates into two times 
younger age 
for the AGB stars in UGC 4483. Again, as in the case of the RGB stars,
we find that the stellar populations seen in UGC 4483 are likely not 
to be older than $\sim$ 2 Gyr, an age which is considerably smaller than 
the age of the universe and which suggests that UGC 4483 is probably a 
relatively young galaxy as compared to other dwarf galaxies. 

\subsection{Surface-brightness and color distributions}

   Another way to study the properties of stellar populations is to consider
their integrated characteristics as given by the surface brightness and color
profiles in different regions in the galaxy. The advantage of this approach 
is that it includes both resolved and unresolved stars. The disadvantage is 
that populations
with different ages contribute to the integrated light and assumptions have to 
be made on the star formation history to derive the age distribution of stars.

\begin{figure}
\figurenum{13}
\epsscale{0.7}
\plotone{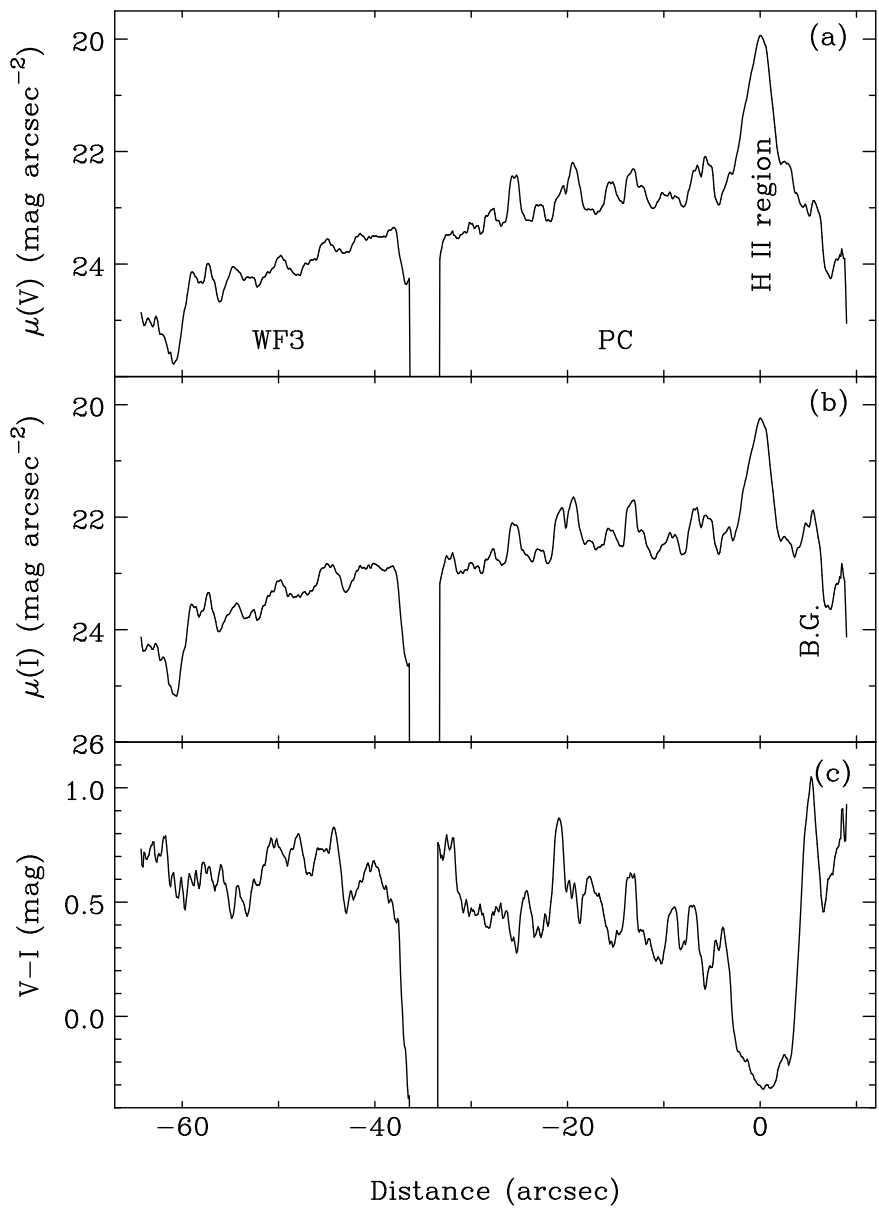}
\caption{$V$ (a) and $I$ (b) surface-brightness and $(V-I)$ color (c)
profiles along the major axis of UGC 4483 (along the diagonals of the PC and 
WF3 frames) in a 2 arcsec wide strip. The distributions are smoothed by a 11 
point box-car. The gap in the profiles at $\sim$ -- 35\arcsec\ is caused by the
lack of signal at the boundaries of the CCD chips (Fig. \ref{Fig1}).
The origin is taken to be at the location of the supergiant H {\sc ii} region
(marked in panel (a)). The red background
galaxy seen in Fig. \ref{Fig2} is marked as B.G. in panel (b).
\label{Fig13}}
\end{figure}

   In Fig. \ref{Fig13}, we show the $V$ (a) and $I$ (b) surface-brightness
and $V-I$ (c) color distributions along the major axis of UGC 4483
in a strip 2\arcsec\ wide centered on the H {\sc ii} region taken to be
the origin.
As before, surface brightnesses and colors have been transformed to the
standard $VI$ photometric system according to the prescriptions
of Holtzman et al. (1995b) and have been corrected for Galactic 
extinction with 
$A_V$ = 0.11 mag (Schlegel et al. 1998). The red background
galaxy to the north of the H {\sc ii} region seen in Fig. \ref{Fig2} is
labeled B.G. in Fig. \ref{Fig13}b. 
Its peak $(V-I)$ color exceeds 1.0 mag (Fig. \ref{Fig13}c).
The gap in Fig. \ref{Fig13} at the distance of $\sim$ --35\arcsec\ is caused
by the lack of signal at the boundaries of the individual CCD chips 
(Fig. \ref{Fig1}). The supergiant H {\sc ii} region has a peak surface 
brightness $\mu$($V$) $\sim$ 20.0 mag arcsec$^{-2}$, or $\sim$ 2 mag
brighter than the surrounding regions. Its profile is broader in $V$ 
(Fig. \ref{Fig13}a) than in $I$ (Fig. \ref{Fig13}b) because of the larger  
contribution of the extended ionized gas emission in $V$. 
Therefore, the region with blue $(V-I)$ color is 
very broad (Fig. \ref{Fig13}).
 The bluest color in this region is $\sim$ --0.3 mag,
bluer than the --0.23 mag color derived for the central ionizing cluster. 
The difference comes from the fact that some ionized gas emission around 
the cluster was subtracted in the derivation of the cluster color, while 
Fig. \ref{Fig13} shows the total stellar and gaseous emission. 

 Fig. \ref{Fig13}c shows an evident $(V-I)$ color gradient. The bluer
colors for distances between --20\arcsec\ and 0 suggest the 
presence of young stellar populations, while the asymptotic value of 
$(V-I)$ $\sim$ 0.7 mag attained for distances between 
--60\arcsec\ and --40\arcsec\
 is characteristic of evolved RGB and AGB stellar populations 
(see {the CMD in Fig. \ref{Fig4}c and} spatial distributions in 
Fig. \ref{Fig9}). This $(V-I)$ color is 
slightly bluer (by $\sim$ 0.1 mag) than the one predicted 
for a $\sim$ 2 Gyr single stellar population by the Padua models with 
$Z$ = 0.0004 and 0.001 
(Tantalo et al. 1996; Girardi et al. 2000;
http://pleiadi.pd.astro.it). A 10 Gyr population would have a predicted
$(V-I)$ $\sim$ 0.9 -- 1.0 mag, which is evidently not the case for UGC 4483.
The combination of a 2 Gyr and older stellar populations, such as produced
by constant star formation between 2 Gyr and 10 Gyr ago, would result in 
a redder color than observed. 

We now compare the asymptotic value of $(V-I)$ with the integrated colors of
the globular clusters considered before in section 3. From 
the catalogue of Galactic globular clusters of Harris (1996) we obtained 
extinction-corrected $(V-I)$ of 0.72, 0.80, 0.84, 0.88, 0.98 
and 1.09 mag respectively  
for the M15, NGC 6397, M2, NGC 6752, NGC 1851 and 47 Tuc 
clusters, arranged in order of increasing metallicity. It can be seen that 
while the observed asymptotic color of 0.7 mag in UGC 4483 
is consistent with that of a 10 Gyr 
population with the metallicity [Fe/H] = --2.17 of M15, the colors of the 
other clusters with higher metallicities are too red.
Therefore, if the metallicity of the stars in UGC 4483 is larger
than that of M15 then we again conclude that their age is younger than that
of globular clusters. It is interesting to note that in spite of the Padua 
10 Gyr theoretical 
isochrones being bluer than those observed for globular clusters, the 
predicted colors of single stellar populations are redder than the integrated 
colors of globular clusters. This is likely due to the steepness of the 
theoretical isochrones which results in too large predicted luminosities 
for the AGB stars.

A relatively young age seems to be
typical of cometary galaxies. 
Thuan, Izotov \& Foltz (1999) found the age of the stellar populations in the 
cometary BCD SBS 1415+437 ($Z_\odot$/21) to be less than $\sim$ 100 Myr.
Although the latter age is probably an underestimate because it was derived 
using instantaneous burst rather than continuous star formation models,
the very blue $V-I$ of $\sim$ 0.4 mag of its extended LSB body argues for 
an age less than 1 Gyr.    
Noeske et al. (2000) found in a detailed study
of two less metal-deficient 
cometary BCDs, Mrk 71($Z_\odot$/14) and Mrk 59 ($Z_\odot$/8),
that their stellar populations are
not older than $\sim$ 3 Gyr in the first one, and $\sim$ 4 Gyr in the second
one. On the basis of this very small sample, there may be a trend of younger 
age with lower metallicity, although a larger sample is needed to really tell. 

Our age estimate of $\sim$ 2 Gyr for UGC 4483 is 
larger than the age of $\la$ 100 Myr suggested by 
Izotov \& Thuan (1999) for the class of very-metal deficient galaxies with 
$Z$ $\la$ 1/20 $Z_\odot$, to which UGC 4483 belongs.  Their argument is based
on the observed constancy of the N/O and C/O ratios with O/H. Izotov \& Thuan
(1999) argued that the constancy meant that carbon and nitrogen in these
metal-deficient galaxies are made by the same massive stars responsible for
oxygen production. In other words, intermediate-mass 
(3$M_\odot$ $\la$ $M$ $\la$ 9$M_\odot$) stars have not had the time to release 
their nucleosynthetic products, which puts an upper limit of $\la$ 40 Myr, the
main-sequence lifetime of a 9$M_\odot$ star. However, there are several 
reasons to believe that the above age upper limit is too low. First,   
the major nitrogen production is not by 8$M_\odot$ stars, but by 
3 -- 4$M_\odot$ 
stars (e.g., Marigo 2001), increasing the time-scale for 
nitrogen enrichment of the interstellar medium by intermediate-mass stars 
to at least to $\sim$ 
150 -- 400 Myr. Second, some additional time 
of $\la$ 1 Gyr (Roy \& Kunth 1995) should be allowed for the mixing of fresh 
nitrogen to the interstellar medium.
 Other possibilities can be 
considered for reconciling the chemical abundance considerations with the age
of UGC 4483. One possible way out
is to note that chemical evolution models of dwarf galaxies follow the time
evolution of local characteristics or assume instantaneous recycling
of the matter. In particular, in closed-box models both oxygen and nitrogen 
are produced at the same place. However, judging from the spatial
distribution of main-sequence stars and AGB stars in UGC 4483 
(Fig. \ref{Fig9}), this assumption may not be hold for this galaxy.
The massive main sequence stars which are the main contributors to the 
interstellar medium of oxygen and probably of primary nitrogen (Izotov \&
Thuan 1999) are located in very compact regions of the galaxy, while the AGB
stars occupy a much larger volume. This may result in a volume dilution of 
the nitrogen produced by AGB stars as compared to the oxygen and
primary nitrogen produced by massive stars, decreasing the relative 
contribution of nitrogen by AGB stars. Other scenarios of nitrogen enrichment 
have been discussed, but they cannot account for the very small dispersion of 
the N/O and C/O ratios in very metal-deficient galaxies 
and the increase in scatter of those ratios in higher-metallicity galaxies
(e.g., Pilyugin 1999; Henry, Edmunds \& K\"oppen 2000).

We note also that the asymptotic $V-I$ of $\sim$ 0.7 of UGC 4483 is the
reddest color of all values derived thus far in our long-term study of 
the stellar populations in extremely metal-deficient BCDs, those with 
$Z$ $\la$ $Z_\odot$/20 (Izotov \& Thuan 1999). The $V-I$ asymptotic colors of 
I Zw 18 ($Z_\odot$/50), SBS 0335--052 ($Z_\odot$/41), SBS 0940+544 
($Z_\odot$/28) and SBS 1415+437 ($Z_\odot$/21) are respectively 0 (Papaderos 
et al. 2001), 0 (Thuan, Izotov \& Lipovetsky 1997), 0.6 (Guseva et al. 2001) 
and 0.4 (Thuan et al. 1999). The colors of I Zw 18 and SBS 0335--052 are 
extremely blue, partly because the contribution of ionized gas 
emission is very 
important, even at large distances from the main star-forming regions.
The ages derived for the oldest stellar populations in the above BCDs are 
all $\la$ 1 Gyr, so that the age of $\sim$ 2 Gyr derived here from 
our CMD analysis of UGC 4483 is consistent with the idea that redder 
$V-I$ colors are caused mainly by age rather than metallicity effects.

\section{Comparison with CMDs of other galaxies}    
      
   There has been several studies of the resolved stellar populations 
and star formation histories in Local Group and other nearby galaxies 
based on {\sl HST} WFPC2 observations during the last few years. 
To put our findings in perspective, we have compared the CMD of UGC 4483 
with those of several dwarf galaxies observed by {\sl HST} with 
the same F555W and F814W filters and for which data are available in the
archives. Our aim was to produce a homogeneous set of CMDs of
dwarf galaxies reduced in the same manner so as to detect
possible general trends.
We have constructed the CMDs of three BCDs, I Zw 18 with an ionized gas 
metallicity $Z_\odot$/50  
(Hunter \& Thronson 1995; Aloisi, Tosi \& Greggio 1999), VII Zw 403 
($Z_\odot$/15, Lynds et al. 1998; Schulte-Ladbeck et al. 1998), UGCA 290
(Crone et al. 2000), 
of the dwarf irregular galaxy NGC 2366 ($Z_\odot$/14, Thuan \& Izotov 2001) 
and of six Local Group dwarf irregular galaxies, 
IC 1613 ($Z_\odot$/13, Cole et 
al. 1999), NGC 6822 ($Z_\odot$/5, Wyder, Hodge \& Zucker 2000), WLM 
($Z_\odot$/15, Dolphin 2000b), Pegasus ($Z_\odot$/10, Gallagher et al. 1998),
Sextans A ($Z_\odot$/27, Dohm-Palmer et al. 1997a, 1997b) and Leo A 
($Z_\odot$/26, Tolstoy et al. 1998). 

\begin{figure}
\figurenum{14}
\epsscale{0.8}
\plotone{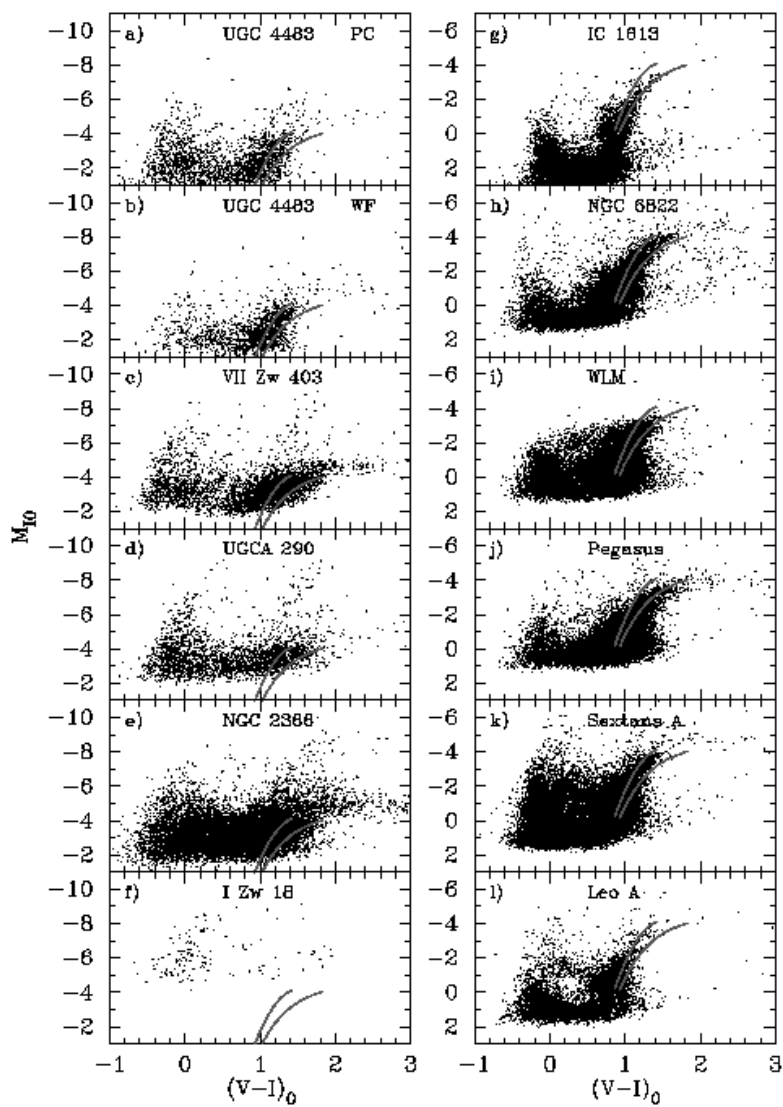}
\caption{CMDs for 5 blue compact dwarf and irregular galaxies outside the Local
Group (panels a to f) and 6 Local Group irregular galaxies (panels g to l)
based on {\sl HST} WFPC2 images. 
For comparison are shown in each panel the isochrones of the globular clusters 
M15 ([Fe/H] = --2.17) (left solid line) and NGC 1851 ([Fe/H] = --1.29) 
(right solid line) (Da Costa \& Armandroff 1990).
\label{Fig14}}
\end{figure}

  In Fig. \ref{Fig14}, we show the CMDs for 
UGC 4483 separately for the PC and WF frames, along with those for the other
10 dwarf galaxies.
The CMD of I Zw 18 was based only on PC data, and that
for Leo A based on WF3 and WF4 data. The CMDs for the other galaxies are
based on all four frames of the WFPC2. We adopt distance moduli $(m-M)$ = 
27.63 for UGC 4483, 28.25 for VII Zw 403 (Lynds
et al. 1998), 29.1 for UGCA 290 (Crone et al. 2000), 27.6 for NGC 2366
(Thuan \& Izotov 2001), 24.27 for IC 1613 (Cole et al. 1999),
23.69 for NGC 6822 (Wyder et al. 2000) 24.88 for WLM (Dolphin 2000b),
24.4 for Pegasus (Gallagher et al. 1998), 25.8 for Sextans A (Dohm-Palmer 
et al. 1997a) and 24.2 for Leo A (Tolstoy et al. 1998). All CMDs have been 
corrected for extinction as quoted in the above mentioned papers. The only
galaxy in Fig. \ref{Fig14} where the RGB population is not detected and hence 
the distance not constrained, is I Zw 18, the BCD with the lowest metallicity
known. We adopt for it a distance modulus $(m-M)$ = 31.0 and correct its
magnitudes and colors for a small Galactic extinction with $E(B-V)$ = 0.032 mag
(Schlegel et al. 1998). The adopted distance modulus for I Zw 18 corresponds 
to a distance of 15.8 Mpc, larger than the redshift distance of
10.8 Mpc adopted by Hunter \& Thronson (1995) and Aloisi et al. 
(1999), and that of 12.6 Mpc used by \"Ostlin (2000). 
The larger distance adopted by us is based on the plausible assumption that the
brightest stars in the CMD of I Zw 18 have absolute $I$ magnitudes in the 
same range as those of the brightest stars in the CMDs of UGC 4483, 
VII Zw 403, UGCA 290 and NGC 2366 which are also undergoing active star 
formation, i.e. between --8 mag and --9 mag. For distances of 10.8 Mpc
and 12.6 Mpc, the absolute magnitudes of the brightest stars in the I Zw 18 
CMD would be 0.83 mag to 0.5 mag fainter, which is hard to
understand. For comparison, we also show in Fig. \ref{Fig14} the observed
isochrones for the two Galactic globular clusters M15 and
NGC 1851 with respective metallicities [Fe/H] = --2.17 and --1.29
(Da Costa \& Armandroff 1990).

   It is evident from Fig. \ref{Fig14}a -- \ref{Fig14}b that the RGB stars in 
UGC 4483 are the bluest as compared to those in other galaxies, supporting the 
conclusion that UGC 4483 may be the youngest among the considered galaxies, 
except for I Zw 18 where the RGB stellar population is not seen. A very 
similar age of 1 -- 2 Gyr has been derived by Tolstoy et al. (1998) 
for the oldest stars of Leo A. They also 
concluded that this galaxy is relatively young. We note however that the AGB 
stars in Leo A (Fig. \ref{Fig14}l) are fainter than those in UGC 4483,
implying that they have a larger age or alternatively, a larger
metallicity. In other Local Group galaxies 
(Fig. \ref{Fig14}g -- \ref{Fig14}k), red giant clump and/or horizontal-branch
stars and RR Lyr variables imply the presence of a several Gyr old stellar 
population. The detection of such stars is not possible with {\sl HST} 
for galaxies beyond the Local Group. 
However, comparing the
CMD of UGC 4483 (Fig. \ref{Fig14}a -- \ref{Fig14}b) with those
of the dwarf galaxies VII Zw 403, UGCA 290 and NGC 2366 
(Fig. \ref{Fig14}c -- \ref{Fig14}e), we conclude that the RGB and AGB
stars in the latter galaxies are likely to be older, 
the RGB stars because their mean color is redder and the 
AGB stars because their mean luminosity is lower.
We have shown in detail the CMDs 
of VII Zw 403 and UGCA 290 respectively in Figures 6c and 
6d. It can be clearly seen that the RGB tip in UGC 4483 is bluer by $\sim$
0.2 mag and its AGB stars are brighter by $\sim$ 0.3 mag than their 
counterparts in VII Zw 403 and UGCA 290. According to Bertelli et al. (1994),
this would mean that the AGB stars in UGC 4483 are about 2 times younger than 
those in VII Zw 403 and UGCA 290. 
We cannot exclude however the possibility that some of these differences are 
also caused by metallicity effects.
Among this group, VII Zw 403 has the largest age and/or stellar 
metallicity because its AGB stars are the least luminous.

\begin{figure}
\figurenum{15}
\epsscale{0.8}
\plotone{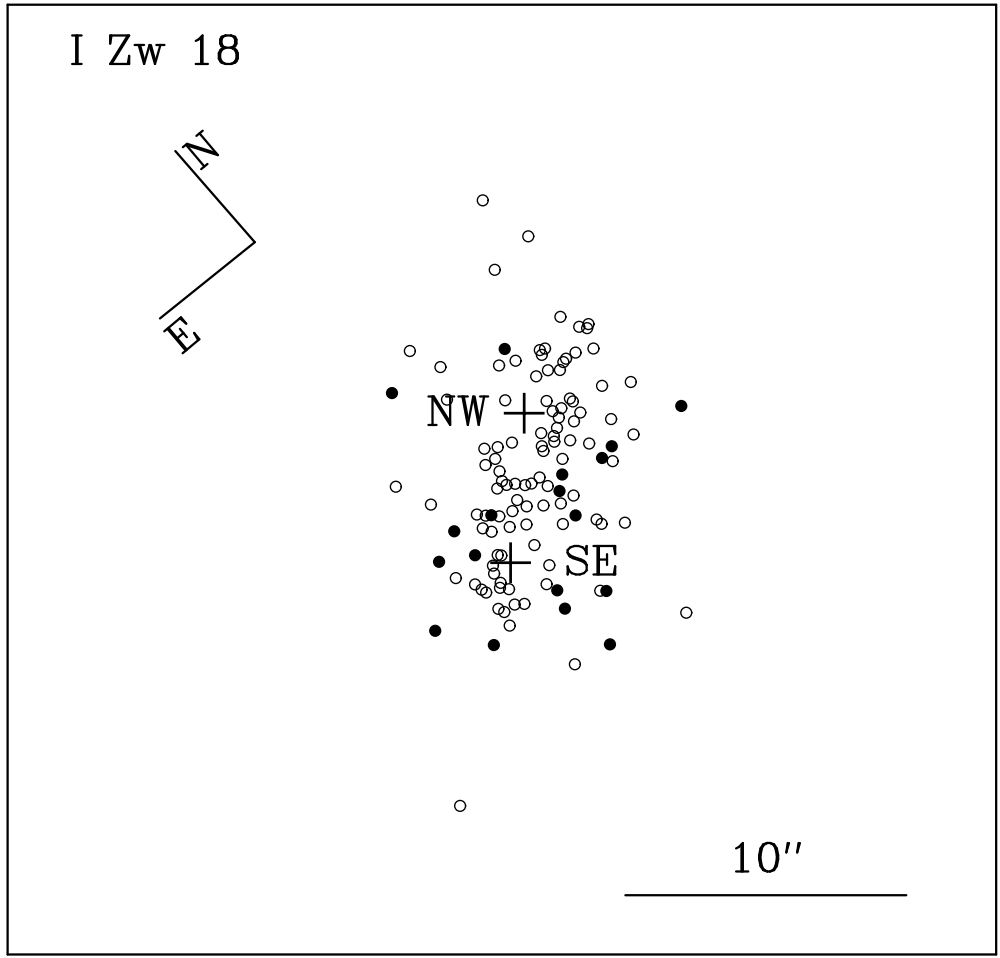}
\caption{Spatial distribution of evolved red ($(V-I)$ $\geq$ 1.0 mag) AGB 
stars (filled circles) and younger bluer ($(V-I)$ $<$ 1.0 mag) stars 
(open circles) in the BCD I Zw 18. The NW and SE star-forming regions
are marked by crosses. Note the relatively compact distribution of the AGB 
stars around the SE region, as compared to their more spread-out distribution
in other dwarf galaxies (see Fig. \ref{Fig9})
\label{Fig15}}
\end{figure}

With our adopted larger distance of 15.8 Mpc, an appreciable number of 
red stars, most likely AGB stars, 
with absolute $I$ 
magnitude of $\sim$ --6 mag or brigher can be seen in the CMD of I Zw 18.
They are $\sim$ 1 mag brighter than in UGC 4483. These stars are 
also bluer than 
in other galaxies, as expected from the low metallicity of I Zw 18.
They were first detected and analyzed by Aloisi et al. (1999) who
concluded that they have ages of at least 1 Gyr. They are
also present in the {\sl HST} NICMOS near-infrared CMD of I Zw 18 by 
\"Ostlin (2000) who derived ages of up to 5 Gyr using Padua theoretical 
isochrones with $Z$ = 0.0004 (Bertelli et al. 1994). Given the uncertainties 
of these isochrones as discussed in section 3, these ages are probably
overestimates and are not likely to exceed 1 Gyr.  Support for a relatively 
small age of the AGB stars in I Zw 18 comes from their spatial distribution 
shown in Fig. \ref{Fig15}. Filled circles denote red stars with 
$(V-I)$ $\geq$ 1.0 mag and open circles all other stars detected in the PC 
image of I Zw 18. The locations of the bright star-forming NW and SE regions 
are shown by crosses. Aloisi et al. (1999) first noted that the red AGB 
stars are located in the south-eastern part of the galaxy. We confirm this 
result. We also note that these AGB stars are detected in the same crowded 
regions where bluer and younger stars are present. Such a distribution is very
different from that seen in UGC 4483, VII Zw 403 (Lynds et al. 1998; 
Schulte-Ladbeck et al. 1998),  
UGCA 290 (Crone et al. 2000) and NGC 2366 (Thuan \& Izotov 2001) where the
RGB and AGB stars are spread out over significantly larger areas as compared to
the younger stars. This difference in spatial distribution may be an 
indication that AGB stars in I Zw 18 have had no time to diffuse far away 
from their origin and hence may be significantly younger than 1 Gyr. The 
compact distribution of the AGB stars and their conspicuous absence at larger 
distances in non-crowded regions where they can be more easily detected, 
also argue 
against the idea that I Zw 18 possesses an extended low-surface-brightness 
disk of red stars (Legrand 2000; Kunth \& \"Ostlin 2000).
Deeper images (by 1--2 mag) of I Zw 18 are required to check 
for the presence or absence of RGB stars in this BCD and settle the 
controversy on its age. This will be possible with the future  
Advanced Camera for Surveys (ADS) to be installed on {\sl HST}.

   In conclusion, {\sl HST} observations of nearby galaxies
with resolved stellar populations have not revealed ones as young as
$\la$ 100 Myr as suggested by Izotov \& Thuan (1999) from chemical arguments. 
However, there is evidence that suggests that some nearby dwarf galaxies, like 
I Zw 18, SBS 0335--052
 and UGC 4483 have ages which are significantly smaller than the age 
of the Universe ($\la$ 2 Gyr), and in that sense, can be considered as
young galaxies.

\section{Summary}

   We present a photometric study of the resolved stellar populations in
the nearby cometary blue compact dwarf galaxy UGC 4483 based on 
{\sl Hubble Space Telescope} WFPC2 $V$ and $I$ images.
The analysis of the color-magnitude diagram (CMD) for this galaxy have led
us to the following conclusions:

   1. The CMD of UGC 4483 is populated by the stars with different ages
including the youngest hydrogen-burning main-sequence stars, evolved massive 
stars with core helium burning (blue-loop stars and red supergiants), 
and older red giants and asymptotic giant helium-shell burning stars. Hence, 
there was continuous star formation in UGC 4483 during a long period, 
for at least the last 2 Gyr.

   2. From the observed $I$ magnitude of the tip of the red giant branch (TRGB)
of 23.65 $\pm$ 0.10 mag, we derive a distance modulus $(m-M)$ = 
27.63 $\pm$ 0.11 corresponding to a distance of 3.4 $\pm$ 0.2 Mpc and 
suggesting that UGC 4483 is a member of the M81 group of galaxies.

   3. The $(V-I)$ color of the red giant stars in UGC 4483 is comparatively
very blue, with a TRGB $(V-I)$ = 1.3 mag implying a very low metallicity 
([Fe/H] $\sim$ --2.4) and/or a young age of the RGB stars. We argue for a
young age of $\sim$ 2 Gyr.

   4. The AGB stars in UGC 4483 are significantly brighter than those observed
in other dwarf galaxies with similar metallicities. They are, for example,
$\sim$ 0.3 mag brighter than those seen in the BCDs VII Zw 403 and UGCA 290. 
Assuming the stars have the same metallicity as the ionized gas, the higher 
luminosities suggest that the AGB stars in UGC 4483 are about a factor of 
2 younger than those in VII Zw 403 and UGCA 290.

   5. We do not find evidence for stellar populations older than 2 Gyr in
UGC 4483, although our data is not deep enough to detect the indicators of 
older stellar populations such as red giant clump and horizontal-branch stars. 
 Comparison of the stellar populations of UGC 4483 with those in other nearby
dwarf galaxies suggests that it is likely to be one of the youngest among the
BCDs and dwarf irregular galaxies with detected RGB and AGB stars.
Alternatively, the differences in the CMD of UGC 4483  
can also be explained by an abnormally low metallicity of the stars in the 
BCD, [Fe/H] = --2.4, comparable to the most metal-deficient globular cluster 
stars.

\acknowledgments
Y.I.I. and T.X.T. are grateful for the partial financial support of NSF 
grant AST-96-16863. T.X.T. has also been supported by HST-GO-08769.01-A grant.
Y.I.I. thanks the hospitality of the Astronomy Department of the University of 
Virginia. We thank M\'arcio Catelan and Robert Rood for useful discussions.

\clearpage

%
\begin{deluxetable}{lrrcrr}
\tablenum{1}
\tablecolumns{3}
\tablewidth{0pt}
\tablecaption{Photometry completeness\tablenotemark{a}
\label{Tab1}}
\tablehead{ Mag. & \multicolumn{2}{c}{PC}&& \multicolumn{2}{c}{WF4} \\
\cline{2-3} \cline{5-6}
 & \multicolumn{1}{c}{F555W} & \multicolumn{1}{c}{F814W} &
& \multicolumn{1}{c}{F555W} & \multicolumn{1}{c}{F814W} }
\startdata
23 -- 24&100.0&100.0&
&100.0&100.0 \\
24 -- 25& 96.4 $\pm$ \,~5.0& 91.3 $\pm$  \,~7.1&
& 93.3 $\pm$ 14.9& 84.4 $\pm$ 14.9 \\
25 -- 26& 80.6 $\pm$ \,~9.0& 78.5 $\pm$ 12.4&
& 80.0 $\pm$ \,~7.5& 72.4 $\pm$ \,~9.2 \\
26 -- 27& 69.5 $\pm$ \,~3.5& 54.4 $\pm$ \,~7.5&
& 66.9 $\pm$ 10.4& 44.8 $\pm$  \,~8.4 \\
27 -- 28& 40.5 $\pm$  \,~5.2& 15.1 $\pm$ \,~7.2&
& 44.2 $\pm$ 12.1&  6.7 $\pm$  \,~9.4 \\
28 -- 29& 11.2 $\pm$  \,~6.6& \nodata~~~~ &
& 12.2 $\pm$ \,~6.0&  \nodata~~~~ \\
\enddata
\tablenotetext{a}{expressed in percentage of recovered stars.}
\end{deluxetable}

\clearpage


\clearpage


\clearpage


\clearpage


\clearpage


\clearpage


\clearpage


\clearpage


\clearpage


\clearpage


\clearpage


\clearpage


\clearpage


\clearpage


\clearpage


\end{document}